\documentclass{elsart}
\usepackage{amsfonts}\usepackage{amsbsy}
\usepackage{mathrsfs}
\usepackage{graphicx}
\def\cal{\mathcal}
\let\sectiontmp\section\let\subsectiontmp\subsection \let\appendixtmp\appendix
\def\section{\setcounter{equation}{0}\sectiontmp}
\def\subsection{\subsectiontmp}
\def\theequation{\arabic{section}.\arabic{equation}}
\def\appendix{\def\theequation{\Alph{section}.\arabic{equation}}\appendixtmp}
\def\ssp{\makebox(0,0)
    {\thinlines\put(-.1,0){\line(1,0){.2}}\put(0,-.1){\line(0,0){.2}}}}
\def\ssm{\makebox(0,0){\put(-.1,0){\thinlines\line(1,0){.2}}}}

\def\propagatorR#1#2#3#4{\begin{picture}(1.4,.8)#3
    \if#4r
    \put(1.3,.1){\circle{.2}}\put(1.1,.3){$R$}\else
    \put(.1,.1){\circle{.2}}\put(-0.1,.3){$R$}\fi
    \put(1.3,0.1){\vector(-1,0){.8}}\put(.1,0.1){\line(1,0){.6}}
    \put(.2,-.15){\makebox(0,0){#1}}
    \put(1.2,-.15){\makebox(0,0){#2}}\end{picture}}
\def\propagatorA#1#2#3#4{\begin{picture}(1.4,.6)#3
    \if#4r
    \put(1.3,.1){\circle{.2}}\put(1.1,.3){$R$}\else
    \put(.1,.1){\circle{.2}}\put(-0.1,.3){$R$}\fi
    \put(0.1,0.1){\vector(1,0){.8}}\put(.7,0.1){\line(1,0){.6}}
    \put(.2,-.15){\makebox(0,0){#1}}
    \put(1.2,-.15){\makebox(0,0){#2}}\end{picture}}
\def\self#1#2#3{\begin{picture}(2.2,.8)\thicklines
    \put(2.3,0.1){\vector(-1,0){1.3}}\put(-.1,0.1){\line(1,0){1.3}}
    \put(1.1,0.1){\oval(2,1)}
    \put(0,-0.15){\makebox(0,0){#1}}\put(2.2,-0.15){\makebox(0,0){#2}}
    \if#3R {\put(0.1,0.1){\circle{.2}}}\else
    \if#3A {\put(2.1,0.1){\circle{.2}}}\else
    {\put(1.1,0.1){\makebox(0,0){#3}}}
    \fi\fi\end{picture}}

\def\contourxy{\unitlength=0.8cm
\begin{picture}(17,1)\thicklines
\contour
\put(14.5,-.5){\makebox(0,0){$\infty$}}
\put(11,-.5){\makebox(0,0){$t_x^+$}}
\put(8,1.8){\makebox(0,0){$t_y^-$}}
\put(11,0){\circle*{.2}}
\put(8,1){\circle*{.2}}
\end{picture}}

\def\contour{\thicklines
\put(1,-.5){\makebox(0,0){$t_{0}$}}
\put(16.5,.5){\makebox(0,0){$t$}}
\put(0,.5){\vector(1,0){16}}
\put(1,1.){\line(1,0){13}}\put(14,.5){\oval(1,1)[br]}
\put(14,0){\vector(-1,0){13}}\put(14,.5){\oval(1,1)[tr]}}

\linethickness{1.0pt}

\def\GG{\begin{picture}(24,0)
\put(22,2){\thinlines\vector(-
1,0){12}}\put(2,2){\thinlines\line(1,0){10}}
     \end{picture}}
\def\GGfull{\begin{picture}(24,0)
\put(22,2){\thicklines\vector(-
1,0){12}}\put(2,2){\thicklines\line(1,0){10}}
     \end{picture}}
\def\GGxy{\begin{picture}(24,0)
\put(22,2){\thinlines\vector(-
1,0){12}}\put(2,2){\thinlines\line(1,0){10}}
\put(4,-3){\makebox(0,0){$x$}}\put(20,-3){\makebox(0,0){$y$}}
     \end{picture}}
\def\GGfullxy{\begin{picture}(24,0)
\put(22,2){\thicklines\vector(-
1,0){12}}\put(2,2){\thicklines\line(1,0){10}}
\put(4,-3){\makebox(0,0){$x$}}\put(20,-3){\makebox(0,0){$y$}}
     \end{picture}}
\def\Dysonself{\begin{picture}(60,0)
\put(0,2){\thicklines
     \put(0,0){\thinlines\line(1,0){10}}
        \put(20,0){\thinlines\vector(-1,0){12}}
     \put(60,0){\vector(-1,0){12}}\put(40,0){\line(1,0){10}}
     \put(30,0){\oval(20,15)}
     \put(30,0){\makebox(0,0){$-\ii\Se$}}}
     \end{picture}}

\def\gsand4{\begin{picture}(38,0)
\put(2,2){
     \thicklines
     \put(2,0){\circle*{2.00}}
     \put(32,0){\circle*{2.00}}
     \bezier{216}(2,0)(17,-10)(32,0)
     \bezier{216}(2,0)(17,+10)(32,0)
     \bezier{284}(2,0)(17,-25)(32,0)
     \bezier{284}(2,0)(17,+25)(32,0)}
     \end{picture}}
\def\ssand3{\begin{picture}(38,0)
\put(2,2){
     \thicklines
     \put(2,0){\line(1,0){30}}
     \put(2,0){\circle*{4.00}}
     \put(32,0){\circle*{4.00}}
     \bezier{284}(2,0)(17,-25)(32,0)
     \bezier{284}(2,0)(17,+25)(32,0)}
     \end{picture}}
\def\sphisand2phi{\begin{picture}(42,0)
\put(2,2){
     \thicklines
     \put(2,0){
     \put(2,0){\line(1,0){8}}
     \put(24,0){\line(1,0){8}}
     \put(10,0){\makebox(0,0){$\bigotimes$}}
     \put(24,0){\makebox(0,0){$\bigotimes$}}
     \put(2,0){\circle*{4.00}}
     \put(32,0){\circle*{4.00}}
     \bezier{284}(2,0)(16,-25)(32,0)
     \bezier{284}(2,0)(16,+25)(32,0)}}
     \end{picture}}
\def\gphisand3phi{\begin{picture}(53,0)
\put(2,2){
     \thicklines
     \put(3,0){\line(1,0){44}}
     \put(3,0){\makebox(0,0){$\bigotimes$}}
     \put(47,0){\makebox(0,0){$\bigotimes$}}
     \put(8,0){
     \put(2,0){\circle*{2.00}}
     \put(32,0){\circle*{2.00}}
     \bezier{284}(2,0)(16,-25)(32,0)
     \bezier{284}(2,0)(16,+25)(32,0)}}
     \end{picture}}

\def\DSa{\begin{picture}(0,0)\thicklines
\put(0,0){\oval(1.5,1)}
\put(0,0){\makebox(0,0){$-\ii\Sa$}}\end{picture}}
\def\GlnG0Sa{
\begin{picture}(4.3,1.5)
\put(0.5,.1){
\put(.5,0){\DSa}\put(3,0){\DSa}\put(1.75,1){\DSa}
\put(1.75,-1){\makebox(0,0){\dots\dots}}
\put(1,.5){\oval(1,1)[lt]}\put(2.5,.5){\oval(1,1)[rt]}
\put(1,-.5){\oval(1,1)[lb]}\put(2.5,-.5){\oval(1,1)[rb]}}
\end{picture}}
\def\GGaSa{
\begin{picture}(2.5,2.)
\thicklines\put(0.5,.1){
\put(.5,0){\DSa}\put(2,-.5){\line(0,1){1}}
\put(1.,1){\line(1,0){0.5}}\put(1.,-1){\line(1,0){0.5}}
\put(1,.5){\oval(1,1)[lt]}\put(1.5,.5){\oval(1,1)[rt]}
\put(1,-.5){\oval(1,1)[lb]}\put(1.5,-.5){\oval(1,1)[rb]}}
\end{picture}}
\def\vhight#1{\vphantom{\left(\begin{picture}(0,#1)\end{picture}\right)}
}
\def\Dclosed#1#2{
\begin{picture}(2,.8)\put(0,.1){#2
\put(1,0){\circle{1.414}}
\put(1,-.707){\line(0,1){1.414}}\put(.293,0){\line(1,0){1.414}}
\put(0.5,-0.5){\line(0,1){1}}\put(1.5,-0.5){\line(0,1){1}}
\put(0.5,-0.5){\line(1,0){1}}\put(0.5,0.5){\line(1,0){1}}}
\put(1.5,-.8){$#1$}
\end{picture}}
\def\loopa{
\unitlength 1.00mm
\thicklines
\begin{picture}(10.00,8.93)
\bezier{28}(2.00,5.00)(2.13,8.33)(6.00,8.93)
\bezier{28}(2.00,4.87)(2.13,1.53)(6.00,0.93)
\bezier{28}(10.00,5.00)(9.87,8.33)(6.00,8.93)
\bezier{28}(10.00,4.87)(9.87,1.53)(6.00,0.93)
\bezier{28}(2.00,-3.00)(2.13,0.33)(6.00,0.93)
\bezier{28}(2.00,-3.13)(2.13,-6.47)(6.00,-7.07)
\bezier{28}(10.00,-3.00)(9.87,0.33)(6.00,0.93)
\bezier{28}(10.00,-3.13)(9.87,-6.47)(6.00,-7.07)
\put(6.00,1.00){\circle*{1.00}}
\put(6.40,8.80){\vector(1,0){0.2}}
\put(5.53,8.73){\line(1,0){0.87}}
\put(5.53,-6.93){\vector(-1,0){0.2}}
\put(6.47,-7.00){\line(-1,0){0.93}}
\end{picture}}
\def\loopb#1{
\unitlength 1.00mm
\thicklines
\begin{picture}(17.50,13.00)
\put(2.00,1.00){\circle*{1.00}}
\put(17.00,1.00){\circle*{1.00}}
\bezier{68}(2.00,1.00)(9.50,5.00)(17.00,1.00)
\bezier{68}(2.00,1.00)(9.50,-3.00)(17.00,1.00)
\bezier{112}(2.00,1.00)(9.50,13.00)(17.00,1.00)
\bezier{112}(2.00,1.00)(9.50,-11.00)(17.00,1.00)
\put(9.93,-1.07){\vector(1,0){0.2}}
\put(9.00,-1.07){\line(1,0){0.93}}
\put(9.93,2.93){\vector(1,0){0.2}}
\put(8.93,2.93){\line(1,0){1.00}}
\put(9.13,6.93){\vector(-1,0){0.2}}
\put(10.07,6.93){\line(-1,0){0.93}}
\put(9.20,-5.00){\vector(-1,0){0.2}}
\put(10.07,-5.07){\line(-1,0){0.87}}
\if#1c \multiput(9.5,-8.)(0,6.5){3}{\line(0,1){5}}\fi
\end{picture}}
\def\loopc#1{ 
\unitlength 1.00mm
\thicklines
\begin{picture}(18.54,9.47)
\put(2.00,-5.00){\circle*{0.93}}
\put(18.00,-5.00){\circle*{1.07}}
\put(10.00,9.00){\circle*{0.94}}
\bezier{72}(2.00,-5.00)(10.00,-1.00)(18.00,-5.00)
\bezier{72}(2.00,-5.00)(10.00,-9.00)(18.00,-5.00)
\bezier{80}(2.00,-5.00)(2.00,6.00)(10.00,9.00)
\bezier{72}(2.00,-5.00)(9.07,-0.80)(10.00,9.00)
\bezier{72}(18.00,-5.00)(11.07,-0.93)(10.00,9.00)
\bezier{76}(18.00,-5.00)(18.13,5.07)(10.00,9.00)
\put(9.93,-7.00){\vector(-1,0){0.2}}
\put(11.00,-7.00){\line(-1,0){1.07}}
\put(3.80,3.27){\vector(1,2){0.2}}
\multiput(3.40,2.47)(0.10,0.20){4}{\line(0,1){0.20}}
\put(16.73,2.33){\vector(2,-3){0.2}}
\multiput(16.20,3.13)(0.11,-0.16){5}{\line(0,-1){0.16}}
\put(10.07,-3.00){\vector(1,0){0.2}}
\put(9.07,-3.00){\line(1,0){1.00}}
\put(12.53,0.60){\vector(-2,3){0.2}}
\multiput(13.00,-0.20)(-0.12,0.20){4}{\line(0,1){0.20}}
\put(7.60,0.67){\vector(-2,-3){0.2}}
\multiput(8.00,1.33)(-0.10,-0.17){4}{\line(0,-1){0.17}}
\if#1c \multiput(8.5,-9.5)(3.5,5.25){3}{\thinlines\line(2,3){2.5}}\fi
\end{picture}}
\def\loopl{
\unitlength 1.00mm
\thicklines
\begin{picture}(18.07,7.00)
\put(2.00,1.00){\circle*{1.00}}
\put(17.00,1.00){\circle*{1.00}}
\bezier{76}(2.00,1.00)(9.47,7.00)(17.00,1.00)
\bezier{76}(2.00,1.00)(9.60,-5.00)(17.00,1.00)
\put(10.00,3.93){\vector(1,0){0.2}}
\put(9.00,3.93){\line(1,0){1.00}}
\put(9.07,-2.00){\vector(-1,0){0.2}}
\put(10.00,-2.00){\line(-1,0){0.93}}
\put(1.00,-2.00){\makebox(0,0)[cc]{$y^k$}}
\put(18.07,-2.00){\makebox(0,0)[cc]{$x^j$}}
\end{picture}}
\def\pia{
\unitlength 1.00mm
\thicklines
\begin{picture}(10.00,8.93)
\bezier{28}(2.00,5.00)(2.13,8.33)(6.00,8.93)
\bezier{28}(2.00,4.87)(2.13,1.53)(6.00,0.93)
\bezier{28}(10.00,5.00)(9.87,8.33)(6.00,8.93)
\bezier{28}(10.00,4.87)(9.87,1.53)(6.00,0.93)
\put(6.00,1.00){\circle*{1.00}}
\put(3.93,1.00){\vector(1,0){0.2}}
\put(2.93,1.00){\line(1,0){1.00}}
\put(6.60,8.80){\vector(1,0){0.2}}
\put(5.60,8.80){\line(1,0){1.00}}
\put(9.83,1.00){\vector(1,0){0.2}}
\put(3.90,1.00){\line(1,0){5.93}}
\end{picture}}
\def\pib{
\unitlength 1.00mm
\thicklines
\begin{picture}(19.78,13.00)
\put(2.00,1.00){\circle*{1.00}}
\put(17.00,1.00){\circle*{1.00}}
\bezier{112}(2.00,1.00)(9.50,13.00)(17.00,1.00)
\put(9.47,1.00){\vector(1,0){0.2}}
\bezier{28}(2.00,1.00)(9.47,1.00)(9.47,1.00)
\put(9.47,6.93){\vector(-1,0){0.2}}
\put(10.07,6.93){\line(-1,0){0.60}}
\bezier{76}(2.00,1.00)(9.60,6.47)(17.00,1.00)
\put(10.00,3.60){\vector(1,0){0.2}}
\put(9.00,3.60){\line(1,0){1.00}}
\bezier{8}(-0.07,1.00)(1.93,1.00)(1.93,1.00)
\put(0.60,1.07){\vector(1,0){0.2}}
\put(-0.00,1.07){\line(1,0){0.60}}
\put(19.78,1.00){\vector(1,0){0.2}}
\put(9.44,1.00){\line(1,0){10.33}}
\end{picture}}
\def\pic{
\unitlength 1.00mm
\thicklines
\begin{picture}(21.78,15.60)
\put(2.07,1.13){\circle*{0.93}}
\put(18.07,1.13){\circle*{1.07}}
\put(10.07,15.13){\circle*{0.94}}
\bezier{80}(2.07,1.13)(2.07,12.13)(10.07,15.13)
\bezier{72}(2.07,1.13)(9.14,5.33)(10.07,15.13)
\bezier{72}(18.07,1.13)(11.14,5.20)(10.07,15.13)
\bezier{76}(18.07,1.13)(18.20,11.20)(10.07,15.13)
\put(3.87,9.40){\vector(1,2){0.2}}
\multiput(3.47,8.60)(0.10,0.20){4}{\line(0,1){0.20}}
\put(16.80,8.46){\vector(2,-3){0.2}}
\multiput(16.27,9.26)(0.11,-0.16){5}{\line(0,-1){0.16}}
\put(12.60,6.73){\vector(-2,3){0.2}}
\multiput(13.07,5.93)(-0.12,0.20){4}{\line(0,1){0.20}}
\put(7.67,6.80){\vector(-2,-3){0.2}}
\multiput(8.07,7.46)(-0.10,-0.17){4}{\line(0,-1){0.17}}
\put(10.07,1.13){\vector(1,0){0.2}}
\bezier{32}(2.07,1.13)(10.07,1.13)(10.07,1.13)
\bezier{8}(-0.00,1.00)(1.00,1.00)(2.00,1.00)
\put(0.73,1.00){\vector(1,0){0.2}}
\put(-0.07,1.00){\line(1,0){0.80}}
\put(21.78,1.11){\vector(1,0){0.2}}
\put(10.22,1.11){\line(1,0){11.56}}
\end{picture}}
\def\pict{
\unitlength 1.00mm
\thicklines
\begin{picture}(21.44,19.00)
\put(2.07,1.13){\circle*{0.93}}
\put(18.07,1.13){\circle*{1.07}}
\put(10.07,15.13){\circle*{0.94}}
\bezier{80}(2.07,1.13)(2.07,12.13)(10.07,15.13)
\bezier{72}(2.07,1.13)(9.14,5.33)(10.07,15.13)
\bezier{72}(18.07,1.13)(11.14,5.20)(10.07,15.13)
\bezier{76}(18.07,1.13)(18.20,11.20)(10.07,15.13)
\put(3.87,9.40){\vector(1,2){0.2}}
\multiput(3.47,8.60)(0.10,0.20){4}{\line(0,1){0.20}}
\put(16.80,8.46){\vector(2,-3){0.2}}
\multiput(16.27,9.26)(0.11,-0.16){5}{\line(0,-1){0.16}}
\put(12.60,6.73){\vector(-2,3){0.2}}
\multiput(13.07,5.93)(-0.12,0.20){4}{\line(0,1){0.20}}
\put(7.67,6.80){\vector(-2,-3){0.2}}
\multiput(8.07,7.46)(-0.10,-0.17){4}{\line(0,-1){0.17}}
\put(10.07,1.13){\vector(1,0){0.2}}
\bezier{32}(2.07,1.13)(10.07,1.13)(10.07,1.13)
\bezier{8}(-0.00,1.00)(1.00,1.00)(2.00,1.00)
\put(0.73,1.00){\vector(1,0){0.2}}
\put(-0.07,1.00){\line(1,0){0.80}}
\put(2.07,-3.07){\makebox(0,0)[cc]{$y_k$}}
\put(18.07,-3.00){\makebox(0,0)[cc]{$x_j$}}
\put(10.00,19.00){\makebox(0,0)[cc]{$z^l$}}
\put(21.44,1.11){\vector(1,0){0.2}}
\put(10.00,1.11){\line(1,0){11.44}}
\end{picture}}
\def\ssp{\makebox(0,0)
  {\thicklines\put(-.1,0){\line(1,0){.2}}\put(0,-.1){\line(0,0){.2}}}}
\def\ssm{\makebox(0,0){\put(-.1,0){\thicklines\line(1,0){.2}}}}

\def\phidecomposition{
     \put(2,0){\thicklines\oval(3.0,1.8)[l]
     \put(-.7,0){\makebox(0,0){$\alpha$}}}
     \put(3,0){\thicklines\oval(3.0,1.8)[r]
     \put(.8,0){\makebox(0,0){$\beta$}}}
     \thicklines
     \put(2,-.9){\line(0,1){1.8}}
     \put(3,-.9){\line(0,1){1.8}}
     \put(2.5,0){\thicklines
     \multiput(0,0.25)(0,0.22){3}{\put(-.5,0){\vector(1,0){1}}
                            \put(-.7,0){\ssp}
                            \put(0.7,0){\ssm} }
     \multiput(0,-.25)(0,-.22){3}{\put(0.5,0){\vector(-1,0){1}}
                            \put(-.7,0){\ssp}
                            \put(0.7,0){\ssm} }}}
\def\wa#1{
\unitlength 0.7mm
\thicklines
\begin{picture}(7,5)\put(2,0){
\put(2.00,1.00){\circle*{1.00}}
\put(2.00,6.00){\makebox(0,0){$ #1 $}}
\bezier{16}(0.54,-0.41)(3.54,2.55)(3.54,2.55)
\bezier{16}(0.54,2.55)(3.50,-0.41)(3.50,-0.41)
\put(1.21,0.22){\vector(1,1){0.2}}
\multiput(0.54,-0.45)(0.11,0.11){6}{\line(0,1){0.11}}
\put(4.61,3.51){\vector(1,1){0.8}}
\multiput(2.32,1.22)(0.11,0.11){20}{\line(0,1){0.11}}
\put(1.25,1.76){\vector(3,-4){0.2}}
\multiput(0.58,2.55)(0.11,-0.13){6}{\line(0,-1){0.13}}
\put(4.54,-1.37){\vector(1,-1){0.8}}
\multiput(2.25,0.76)(0.13,-0.12){18}{\line(1,0){0.13}}
}
\end{picture}}
\def\wb#1{
\unitlength 0.7mm
\thicklines
\begin{picture}(17.51,9.01)\put(4,0){
\put(9.51,-6.49){\circle*{1.00}}
\put(9.51,8.51){\circle*{1.00}}
\put(9.51,-9.49){\makebox(0,0){$ #1 $}}
\put(9.51,11.51){\makebox(0,0){$ #1 $}}
\bezier{68}(9.51,-6.49)(5.51,1.01)(9.51,8.51)
\bezier{68}(9.51,-6.49)(13.51,1.01)(9.51,8.51)
\put(11.58,1.44){\vector(0,1){0.2}}
\put(11.58,0.51){\line(0,1){0.93}}
\put(7.52,0.58){\vector(0,-1){0.2}}
\put(7.52,1.51){\line(0,-1){0.93}}
\bezier{20}(7.00,8.55)(12.04,8.55)(12.04,8.55)
\bezier{20}(7.00,-6.49)(12.04,-6.53)(12.04,-6.53)
\put(8.00,8.51){\vector(1,0){0.2}}
\put(7.00,8.51){\line(1,0){1.00}}
\put(13.44,8.55){\vector(1,0){0.2}}
\put(12.44,8.55){\line(1,0){1.00}}
\put(8.00,-6.58){\vector(1,0){0.2}}
\put(7.00,-6.58){\line(1,0){1.00}}
\put(13.44,-6.58){\vector(1,0){0.2}}
\put(12.44,-6.58){\line(1,0){1.00}}
}
\end{picture}
}
\def\fiqa{
\unitlength 1.00mm
\thicklines
\begin{picture}(47.32,23.09)
\put(2.00,1.00){\circle*{1.00}}
\put(17.00,1.00){\circle*{1.00}}
\bezier{68}(2.00,1.00)(9.50,5.00)(17.00,1.00)
\bezier{68}(2.00,1.00)(9.50,-3.00)(17.00,1.00)
\put(9.93,-1.07){\vector(1,0){0.2}}
\put(9.00,-1.07){\line(1,0){0.93}}
\put(31.93,1.00){\circle*{1.00}}
\bezier{68}(16.93,1.00)(24.43,5.00)(31.93,1.00)
\bezier{68}(16.93,1.00)(24.43,-3.00)(31.93,1.00)
\put(24.86,-1.07){\vector(1,0){0.2}}
\put(23.93,-1.07){\line(1,0){0.93}}
\put(23.93,3.07){\vector(-1,0){0.2}}
\put(24.87,3.07){\line(-1,0){0.94}}
\put(9.07,3.07){\vector(-1,0){0.2}}
\put(10.00,3.07){\line(-1,0){0.93}}
\put(46.82,1.11){\circle*{1.00}}
\bezier{68}(31.82,1.11)(39.32,5.11)(46.82,1.11)
\bezier{68}(31.82,1.11)(39.32,-2.89)(46.82,1.11)
\put(39.75,-0.96){\vector(1,0){0.2}}
\put(38.82,-0.96){\line(1,0){0.93}}
\put(38.82,3.18){\vector(-1,0){0.2}}
\put(39.76,3.18){\line(-1,0){0.94}}
\bezier{252}(2.11,0.98)(24.22,23.09)(46.89,1.20)
\put(24.86,12.04){\vector(1,0){0.2}}
\put(23.93,12.04){\line(1,0){0.93}}
\bezier{252}(2.11,1.02)(24.22,-21.09)(46.89,0.80)
\put(23.93,-10.26){\vector(-1,0){0.2}}
\put(24.87,-10.26){\line(-1,0){0.94}}
\end{picture} }
\def\fipi{
\unitlength 1.10mm
\thicklines
\begin{picture}(17.50,14.91)
\put(2.00,1.00){\circle*{1.00}}
\put(17.00,1.00){\circle*{1.00}}
\put(9.96,7.85){\vector(1,0){0.2}}
\put(8.96,7.85){\line(1,0){1.00}}
\bezier{128}(2.01,0.99)(9.51,14.91)(17.01,0.99)
\bezier{128}(2.01,0.99)(9.51,-12.93)(17.01,0.99)
\put(9.07,-6.00){\vector(-1,0){0.2}}
\put(10.00,-6.00){\line(-1,0){0.93}}
\bezier{8}(2.01,0.99)(2.42,1.99)(3.05,0.99)
\bezier{8}(3.05,0.99)(3.46,-0.01)(4.05,0.99)
\bezier{8}(4.01,0.99)(4.42,1.99)(5.05,0.99)
\bezier{8}(6.01,0.99)(6.42,1.99)(7.05,0.99)
\bezier{8}(8.01,0.99)(8.42,1.99)(9.05,0.99)
\bezier{8}(10.01,0.99)(10.42,1.99)(11.05,0.99)
\bezier{8}(12.01,0.99)(12.42,1.99)(13.05,0.99)
\bezier{8}(14.01,0.99)(14.42,1.99)(15.05,0.99)
\bezier{8}(16.01,0.99)(16.42,1.99)(17.05,0.99)
\bezier{8}(5.05,0.99)(5.46,-0.01)(6.05,0.99)
\bezier{8}(7.05,0.99)(7.46,-0.01)(8.05,0.99)
\bezier{8}(9.05,0.99)(9.46,-0.01)(10.05,0.99)
\bezier{8}(11.05,0.99)(11.46,-0.01)(12.05,0.99)
\bezier{8}(13.05,0.99)(13.46,-0.01)(14.05,0.99)
\bezier{8}(15.05,0.99)(15.46,-0.01)(16.05,0.99)
\end{picture}}
\def\bosprs{
\unitlength 1.00mm
\thinlines
\begin{picture}(17.50,14.91)
\bezier{8}(2.01,0.99)(2.42,1.99)(3.05,0.99)
\bezier{8}(3.05,0.99)(3.46,-0.01)(4.05,0.99)
\bezier{8}(4.01,0.99)(4.42,1.99)(5.05,0.99)
\end{picture}}
\def\bosse{
\unitlength 1.00mm
\thicklines
\begin{picture}(17.50,14.91)
\put(2.00,1.00){\circle*{1.00}}
\put(17.00,1.00){\circle*{1.00}}
\put(9.96,7.85){\vector(1,0){0.2}}
\put(8.96,7.85){\line(1,0){1.00}}
\bezier{128}(2.01,0.99)(9.51,14.91)(17.01,0.99)
\bezier{128}(2.01,0.99)(9.51,-12.93)(17.01,0.99)
\put(9.07,-6.00){\vector(-1,0){0.2}}
\put(10.00,-6.00){\line(-1,0){0.93}}
\end{picture}}
\def\bosprv{
\unitlength 1.00mm
\thinlines
\begin{picture}(17.50,14.91)
\bezier{8}(2.01,0.99)(2.42,1.99)(3.05,0.99)
\bezier{8}(3.05,0.99)(3.46,-0.01)(4.05,0.99)
\bezier{8}(4.01,0.99)(4.42,1.99)(5.05,0.99)
\bezier{8}(6.01,0.99)(6.42,1.99)(7.05,0.99)
\bezier{8}(8.01,0.99)(8.42,1.99)(9.05,0.99)
\bezier{8}(10.01,0.99)(10.42,1.99)(11.05,0.99)
\bezier{8}(12.01,0.99)(12.42,1.99)(13.05,0.99)
\bezier{8}(14.01,0.99)(14.42,1.99)(15.05,0.99)
\bezier{8}(16.01,0.99)(16.42,1.99)(17.05,0.99)
\bezier{8}(5.05,0.99)(5.46,-0.01)(6.05,0.99)
\bezier{8}(7.05,0.99)(7.46,-0.01)(8.05,0.99)
\bezier{8}(9.05,0.99)(9.46,-0.01)(10.05,0.99)
\bezier{8}(11.05,0.99)(11.46,-0.01)(12.05,0.99)
\bezier{8}(13.05,0.99)(13.46,-0.01)(14.05,0.99)
\bezier{8}(15.05,0.99)(15.46,-0.01)(16.05,0.99)
\end{picture}}
\def\bosprf{
\unitlength 1.00mm
\thicklines
\begin{picture}(17.50,14.91)
\bezier{8}(2.01,0.99)(2.42,1.99)(3.05,0.99)
\bezier{8}(3.05,0.99)(3.46,-0.01)(4.05,0.99)
\bezier{8}(4.01,0.99)(4.42,1.99)(5.05,0.99)
\bezier{8}(6.01,0.99)(6.42,1.99)(7.05,0.99)
\bezier{8}(8.01,0.99)(8.42,1.99)(9.05,0.99)
\bezier{8}(10.01,0.99)(10.42,1.99)(11.05,0.99)
\bezier{8}(12.01,0.99)(12.42,1.99)(13.05,0.99)
\bezier{8}(14.01,0.99)(14.42,1.99)(15.05,0.99)
\bezier{8}(16.01,0.99)(16.42,1.99)(17.05,0.99)
\bezier{8}(5.05,0.99)(5.46,-0.01)(6.05,0.99)
\bezier{8}(7.05,0.99)(7.46,-0.01)(8.05,0.99)
\bezier{8}(9.05,0.99)(9.46,-0.01)(10.05,0.99)
\bezier{8}(11.05,0.99)(11.46,-0.01)(12.05,0.99)
\bezier{8}(13.05,0.99)(13.46,-0.01)(14.05,0.99)
\bezier{8}(15.05,0.99)(15.46,-0.01)(16.05,0.99)
\end{picture}}

\newcommand{\di}{{\mathrm d}}
\newcommand{\Tr}{{\mathrm{Tr}}}
\newcommand{\ii}{{\mathrm i}}

\renewcommand{\and}{\quad{\mathrm{and}}\quad}

\renewcommand{\Re}{{\mathrm{Re}}}
\renewcommand{\oint}{\int_{\cal C}}
\renewcommand{\Im}{{\mathrm{Im}}}
\def\Tc{{\cal T}_{\cal C}}

\def\scr#1{\mbox{\scriptsize #1}}
\def\vecs#1{\mbox{\scriptsize \boldmath $#1$}}
\def\vec#1{\mbox{\boldmath $#1$}}
\newcommand{\dpi}[1]{\frac{\di^4 #1}{(2\pi)^4}}                
\newcommand{\Pbr}[1]{\left\{#1\right\}}                    
\newlength{\charwidth}
\def\medhat#1{\settowidth{\charwidth}{$#1\,$}{\makebox[\charwidth]{$\,
 {\widehat{\makebox[2mm]{$#1\,$}}}$}}\vphantom{#1}}
\newcommand{\lap}%
{\raisebox{-0.5ex}{$\stackrel{\scriptstyle <}{\scriptstyle \sim}$}}
\newcommand{\gap}%
{\raisebox{-0.5ex}{$\stackrel{\scriptstyle >}{\scriptstyle \sim}$}}
\def\sgn{\mathrm{sign}}
\def\Gr{G}\def\Se{\Sigma}

\def\Pt{\medhat{\varphi}}\def\Ptd{\Pt^\dagger}

\def\dc{\delta_{\cal C}}
\def\Ga{\Gr}
\def\Sa{\Se}

\def\Lg{{\cal L}}
\def\Lgh{\makebox[3.5mm]{${\widehat{\makebox[2mm]{$\Lg$}}}$}\vphantom{L}}
\def\Lint{\Lgh^{\mbox{\scriptsize int}}}

\def\Hint{\medhat{H}^{\mbox{\scriptsize int}}}

\def\A{A}
\def\Gm{\Gamma}
\def\F{F}                             
\def\Ft{\widetilde{F}}                 
\def\Fd{F}                             
\def\Fdt{\widetilde{F}}                
\def\fd{f}                             

\def\tp{\widetilde{p}}\def\tm{\widetilde{m}}\def\tW{\widetilde{W}} 
\def\Get{\Gamma_{\scr{out}}}   
\def\Gbt{\Get}                                   
\def\Ge{\Gamma_{\scr{in}}}     
\def\Gb{\Ge}              
\def\Loop{L}
\def\Ld{\Gamma_{\scr{out}}}
\def\Ldt{\Gamma_{\scr{in}}}
\def\vu{v}

\def\Do{{\cal D}}

\begin{document}
\begin{frontmatter}
\def\huge{\LARGE}
\title{\vspace*{-15mm}
Resonance Transport  and  Kinetic Entropy} 

\author{Yu. B. Ivanov$^{1,2}$, J. Knoll$^{1}$ 
and D. N. Voskresensky$^{1,3}$} 

\maketitle

\noindent                        
$^{1}${\it Gesellschaft f\"ur Schwerionenforschung mbH, Planckstr. 1,
64291 Darmstadt, Germany}
\\
$^{2}${\it Kurchatov Institute, Kurchatov sq. 1, Moscow 123182,
Russia} 
\\
$^{3}${\it Moscow Institute for Physics and Engineering, 
Kashirskoe sh. 31, Moscow 115409, Russia}                    

\begin{abstract}
  
  Within the real-time formulation of non-equilibrium field theory generalized
  transport equations are derived avoiding the standard quasiparticle
  approximation. They permit to include unstable particles into the transport
  scheme. In order to achieve a self-consistent, conserving and
  thermodynamically consistent description, we generalize the Baym's
  $\Phi$-functional method to genuine non-equilibrium processes.  This scheme
  may be closed at any desired loop order of the diagrams of the functional
  $\Phi$ this way defining a consistent effective theory. By means of a
  first-order gradient approximation the corresponding Kadanoff-Baym equations
  are converted into a set of coupled equations. This set consists of a
  time-irreversible generalized kinetic equation for the slowly varying
  space-time part of the phase-space distributions and a retarded equation,
  which provides the fast micro-scale dynamics represented by the
  four-momentum part of the distributions. Thereby, no constraint to the mass
  shell of the particles is required any further and the corresponding
  spectral mass distributions are treated dynamically. The description
  naturally includes all those quantum features already inherent in the
  corresponding equilibrium limit (Matsubara formalism).  Memory effects
  appearing in collision term diagrams of higher order are discussed.  The
  variational properties of $\Phi$-functional permit to derive a generalized
  expression for the non-equilibrium kinetic entropy flow, which includes
  corrections from fluctuations and mass width effects. In special cases an
  $H$-theorem can be demonstrated implying that the entropy can only increase
  with time.  Memory effects in the kinetic terms provide corrections to the
  kinetic entropy flow that in equilibrium limit recover the famous bosonic
  type $T^3 \ln T$ correction to the specific heat of Fermi liquids like
  Helium-3.

\end{abstract}
\end{frontmatter}

\section{Introduction}

A proper dynamical scheme in terms of a transport concept that deals with
unstable particles, like resonances, i.e. particles with a broad mass width, is
still lacking. Rather ad-hoc recipes are in use that sometimes violate basic
requirements as given by fundamental symmetries and conservation laws,
detailed balance or thermodynamic consistency. The problem of conserving
approximation has first been addressed by Baym and Kadanoff \cite{KadB,Baym}.
They started from an equilibrium description in the imaginary time formalism
and discussed the response to external disturbances.  Baym, in particular,
showed \cite{Baym} that any approximation, in order to be conserving, must be
so-called $\Phi$-derivable. It turned out that the $\Phi$ functional required
is precisely the auxiliary functional introduced by Luttinger and Ward
\cite{Luttinger} (see also ref. \cite{Abrikos}) in connection with the
thermodynamic potential. In nonequilibrium case the problem of conserving
approximations is even more severe than in near-equilibrium linear-response
theory.

The appropriate frame for the description of nonequilibrium processes is the
real-time formalism of quantum field theory, developed by Schwinger, Kadanoff,
Baym and Keldysh \cite{Schw,Kad62,Keld64}.  In ref. \cite{IKV} we have
introduced the generating functional $\Phi$ on the real-time contour. It is
determined by the sum of all closed vacuum skeleton diagrams in terms of full
classical fields and full Green's functions.  All important quantities of a
system (such as the sources of classical fields, self-energies, interaction
energy, etc.)  are derived by variations of the $\Phi$ functional with respect
to its arguments.  The advantage of the $\Phi$ functional is that one may
formulate various approximations in terms of approximate $\Phi$ (so called
$\Phi$-derivable approximations), which preserve the conservation laws related
to global symmetries of the underlying theory and thermodynamic consistency.
Thereby, one may restrict oneself to either few diagrams only or to some
sub-set of diagrams for $\Phi$.  Note that far not every approximation scheme,
e.g., at the level of self-energies, possesses such properties.

Functional variational methods allow to derive the corresponding Dyson
equations on real-time contour. The steps towards generalized kinetic
equations are provided by the Wigner transformation of all two-point functions
and by the first-order gradient approximation. The former formulates all
quantities in terms of phase-space distribution functions in four dimensions,
i.e. as a function of energy and momentum for any space-time coordinate. The
gradient approximation leads to Poisson-bracket expressions which permit a
classical interpretation.  For complicated collision terms, the latter also
give rise to memory effects. The quasiparticle approximation is not required
at any step. Rather one can keep the finite damping width of the particles.
The corresponding dynamical information contained in the spectral functions is
determined by  local retarded equations. We derive closed
expressions for the nonequilibrium energy momentum tensor in the
$\Phi$-derivable scheme, which also include terms arising from the
fluctuations in the system and the finite mass width of the particles.

The second main issue of this paper is the role of entropy.  Although the
entropy is a central quantity in thermodynamics and statistical mechanics,
many problems connected with it, in particular, its description in terms of
Green's functions in the nonequilibrium case, are still open. One can find
related discussions in many textbooks and reviews, e.g., in refs
\cite{Zubarev,Balescu,Wehrl,Klimontovich,Cohen}.  The thermodynamic entropy
has extensively been discussed in the literature at the end of sixties and
beginning of seventies, cf. refs \cite{Riedel,Carneiro,Baym91} and
references therein. The fact that the generalized kinetic equation possesses a
proper thermodynamic limit, does not yet imply that this limit will be
approached during the evolution.  The latter is however ensured, if one can
prove an $H$-theorem for the equations of motion. Using the $\Phi$-derivable
properties we are able to get an expression, which takes
the sense of a nonequilibrium kinetic entropy expressed in terms of Green's
functions and self-energies. It generalizes to nonequilibrium the well known equilibrium
expression obtained in Matsubara formalism within the $\Phi$-functional
scheme.  We derived conditions for an $H$-theorem, which apply in
many cases discussed in transport problems. In addition, the kinetic entropy
expression is extended beyond the quasiparticle limit thus accounting for the
fluctuations. Memory effects lead to further corrections to the entropy flow.
The latter have to be established for each particular case.  In spite of many
attempts so far, $H$-theorem has not yet been proven even for the classical
kinetic equation including triple collision term, cf. ref.  \cite{Cohen}.
Therefore, up to now there exists no appropriate kinetic entropy expression
(derived from generalized kinetic description). Our demonstration of
$H$-theorem for particular cases including the fluctuation parts may be viewed
as step towards this general goal.

The paper is organized as follows. Within the real-time contour formulation of
nonequilibrium dynamics we define a generating functional $\Gamma$ together
with the auxiliary $\Phi$ functional of Baym, cf. ref.  \cite{IKV}, (sect.
\ref{sect-W-Phi}).  Then, for slowly varying generalized distribution
functions we derive a generalized kinetic equation consistent with $\Phi$,
using the gradient approximation (sect.  \ref{sect-Kin-EqT}). Arguments are
given that the generalized distribution functions remain positive during the
time evolution and thus permit a probabilistic interpretation in four-momentum
space at each space-time point. The properties of $\Phi$ determine the
conservation laws in terms of conserved Noether currents and energy-momentum
tensor. Further properties are exploited with the help of the decomposition
rules for the diagrams of $\Phi$ formulated in terms of full Green's functions
\cite{Knoll95}. These rules lead to a multi-process decomposition of a
$\Phi$-derivable collision term, which contains space-time local and nonlocal
(memory) parts (sect.  \ref{Collision Term}).  Time-irreversibility is
discussed in connection with our generalized kinetic description, and
expression for the kinetic entropy flow is derived. In specific cases an
explicit growth of the kinetic entropy with time, i.e. an $H$-theorem (sect.
\ref{H-theorem}) is demonstrated.  The proof once again relies on the
$\Phi$-derivability of the collision term.  Markovian and memory contributions
to the entropy flow are discussed at a specific example. Finally, the proper
thermodynamic limit of the kinetic entropy is demonstrated (sect.
\ref{Ther-entr}).  Some formal details are deferred to Appendices.

To be specific, we concentrate on systems of nonrelativistic particles.
Bosonic mean fields are not treated in this paper; they can however be
included along the lines given in ref. \cite{IKV}. Whenever possible we omit
the index $a$ denoting the different particle species and intrinsic quantum
numbers in a multi-component system.

\newpage
\section{Generating Functionals $\Gamma$ and $\Phi$}\label{sect-W-Phi}

\subsection{$\Phi$ Functional and Dyson's Equation on Real-Time Contour}

We assume the nonequilibrium system to be prepared at some initial time $t_0$
in terms of a given density operator $\medhat{\rho}_0=\sum_{\alpha}
P_{\alpha}\left|\alpha\right>\left<\alpha\right|$, where the
$\left|\alpha\right>$ form a complete set of eigenstates of $\medhat{\rho}_0$.

All observables can be expressed through $n$-point Wightman functions of
Heisenberg operators $\medhat{A}(t_1),\dots ,\medhat{O}(t_n)$ at some later
times
%
\begin{eqnarray}\label{corrfct}
\left<\medhat{O}(t_n)\right. \left.\dots \medhat{B}(t_2)\medhat{A}(t_1)
\right>
&=:&\Tr\; \medhat{O}(t_n)\dots \medhat{B}(t_2)\medhat{A}(t_1)
\medhat{\rho}_0(t_0)
\nonumber\\
&=&
\sum_{\alpha} P_{\alpha}\left<{\alpha}\right|
\medhat{O}(t_n) \dots \medhat{B}(t_2) \medhat{A}(t_1)
\left|{\alpha}\right>.  
\end{eqnarray}
%
Note the fixed operator ordering for Wightman functions.

\parbox[t]{14.5cm}{
\begin{center}\vspace*{1cm}
  \contourxy\\[1cm]
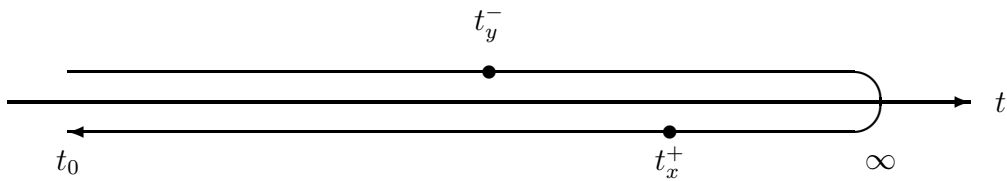
  Figure 1: Closed real-time contour with two external points $x,y$ on
  the contour.
\end{center}}

The nonequilibrium theory can entirely be formulated on a {\em single}
special contour, the {\em closed real-time contour} (see figure 1) with the
time argument running from $t_{0}$ to $\infty$ along {\em time-ordered} branch
and back to $t_{0}$ along {\em anti-time-ordered} branch. The contour Green's
function is defined as the expectation value of contour-ordered products of
operators
%
\begin{eqnarray}\label{Ga1}
\ii\Gr(x,y) =
\left<\Tc\Pt (x)\Ptd(y)\right> ,
\end{eqnarray}
%
where $\Tc$ denotes the special time-ordering operator, which orders the
operators according to a time parameter running along the time contour ${\cal
  C}$. With the aim to come to a self-consistent treatment in terms of
two-point Green's functions, one has to ignore higher order correlations which
would be present in the exact dynamical equation for two-point Green's
functions. This step is achieve by the Wick decomposition which leads to the
Dyson equation now formulated on the real time contour
%
\begin{eqnarray}\label{Dyson}
S_{x}\Gr (x,y) &=& \dc(x,y)+\oint \di z \Se (x,z)\Gr (z,y),\\
\label{S-def}             
S_{x} \Gr^{0} (x,y)
&=&\dc(x,y),
\end{eqnarray}
%
where
\begin{eqnarray}
S_x=
\ii\partial_t+\frac{1}{2m}{\partial_{\vecs x}^2}
\end{eqnarray}
%
in nonrelativistic kinematics.  Here $\dc(x,y)$ is $\delta$-function on the
contour, $\Gr^{0}$ is the free Green's function and $\Se$ is the self-energy,
expressed via one-particle Green's function in the usual way. 
This way one ignores higher order correlation functions, cf. 
refs \cite{Henning,Fauser}.  Then the self-energy itself becomes 
a functional of the Green's functions (\ref{Ga1}) with the 
consequence that the standard Wick decomposition can be applied 
for the self-energy and one thus obtains standard contour 
diagrams, cf. ref.  \cite{IKV}.  In such a diagrammatic 
representation eq. (\ref{Dyson}) is given by \unitlength=.600mm 
%
\begin{eqnarray}
\GGfull&=&\GG + \Dysonself\vphantom{\displaystyle\int_A^B}
\end{eqnarray}
%
with the two-point functions $-\ii\Se\left\{\Gr(x,y)\right\}$, as driving
terms, and the Green's functions
%
\begin{eqnarray}\begin{array}{rlrl}
\ii\Gr^0(x,y)&=\GGxy\quad , \hspace*{1cm}&\ii\Gr(x,y)&=\GGfullxy
\vphantom{\displaystyle\int}\quad.
\end{array}
\end{eqnarray}
%
The arrow always points towards the $\Pt$-field operator in the contour
ordered expression, cf. eq. (\ref{Ga1}).

The physical condition, which justifies to disregard higher order
correlations, is that the typical interaction time $\tau_{\scr{int}}$ for the
change of the correlation functions is much less than the typical relaxation
time $\tau_{\scr{rel}}$, which determines the system evolution.  Describing
the system at times $t-t_0 \gg \tau_{\scr{int}}$, one can neglect initial
correlations, since they are supposed to be dying out at times $\sim
\tau_{\scr{int}}$ in accordance with the Bogolyubov's principle of the
weakening of initial correlations.\footnote{Actually, considering dilute
  systems, Bogolyubov suggested the weakening of all correlations, whereas we
  use a weaker assumption on the weakening of only short-time ($\sim
  \tau_{\scr{int}}$) correlations, cf. ref. \cite{Klimontovich}.  }  This
coarse-graining leads to time-irreversibility, as we will see below.
Alternatively, one also could suppose the initial state to be uncorrelated,
like an equilibrium ideal gas state, which corresponds to an information loss
right from the beginning, cf. refs \cite{Kad62,Dan84}.

In the imaginary-time formalism, Luttinger and Ward \cite{Luttinger} have
shown that the generating functional $\Gamma$ can be expressed in terms of an
auxiliary functional $\Phi$, where $\Phi$ is solely given in terms of full,
i.e. resummed, propagators. Following ref. \cite{IKV}, it can be generalized
to the real-time case with the following diagrammatic representation
\unitlength=.8cm
%
\begin{eqnarray}\label{keediag}
&&\ii\Gamma\left\{\Gr , \lambda \right\} = \ii
\Gamma^0\left\{\Gr^0\right\}  
\nonumber 
\\
&&\hspace*{5mm}
+
\left\{\vhight{1.6}\right.
\underbrace{\sum_{n_\Se}\vhight{1.6}\frac{1}{n_\Se}\GlnG0Sa}
_{\displaystyle \pm \ln\left(1-\odot\Gr^{0}\odot\Se\right)}
\underbrace{-\vhight{1.6}\GGaSa}
_{\displaystyle \pm \odot\Gr\odot\Se\vphantom{\left(\Ga^{0}\right)}}
\left.\vhight{1.6}\right\}
\underbrace{+\vhight{1.6}\sum_{n_\lambda}\frac{1}{n_\lambda}
\Dclosed{c2}{\thicklines}}
_{\displaystyle\vphantom{\left(\Ga^{0}\right)} +\ii\Phi\left\{\Gr , \lambda \right\}}.
\end{eqnarray}
%
Here upper signs relate to fermion quantities, whereas lower signs, to boson
ones. For the sake of mathematical convenience, cf. ref. \cite{IKV}, a scaling
parameter $\lambda$ is introduced in each interaction vertex. For the physical
case, $\lambda = 1$.  The integer $n_\Se$ counts the number of self-energy
insertions in the ring diagrams. For the closed diagrams of $\Phi$ the value
$n_\lambda$ counts the number of vertices building up each such diagram.  Due
to the global factor $1/n_\lambda$ the set of $\Phi$-diagrams is not resumable
in the standard diagrammatic sense. For nonrelativistic instantaneous
two-body interaction each interaction counts as two vertices, cf. Appendix
\ref{diagrules}. The term $\Gamma^0$ solely depends on the free propagator
$\Gr^0$.  The diagrams contributing to $\Phi$ are given in terms of full
propagators $\Gr$ (thick lines). These diagrams are {\em two-particle
  irreducible} (label $c2$), i.e. they cannot be decomposed into two pieces by
cutting two propagator lines.  The latter property matches diagrammatic rules
for the resummed self-energy $\Se(x,y)$, which results from functional
variation of $\Phi$ with respect to the propagator $\Gr(y,x)$, i.e.
\begin{eqnarray}\label{varphi}
-\ii 
\Se =\mp \delta \ii \Phi / \delta \ii \Gr  . 
\end{eqnarray}
Relation (\ref{varphi}) directly follows from the stationarity condition 
of $\Gamma$ (\ref{keediag}) with
respect to variations in the full propagator $\Gr$ on the contour 
%
\begin{eqnarray}
\label{varG/phi}
\delta \Gamma \left\{\Gr , \lambda \right\}/ \delta \Gr = 0,
\end{eqnarray}
%
which also provides the Dyson's equation (\ref{Dyson}). In graphical terms,
the variation (\ref{varphi}) is realized by opening a propagator line in any
diagram of $\Phi$.  The resulting set of thus opened diagrams must then be
that of proper skeleton diagrams of $\Se$ in terms of {\em full propagators},
i.e.  void of any self-energy insertions.

Subsequently we will not use the contour representation but rather its
decomposition into the two branches the time-ordered and the anti-time ordered
branch using a $\{-+\}$ matrix notation as explained in Appendix
\ref{Contour}, cf. also refs \cite{Lif81,IKV}.

\subsection{ $\Phi$-Derivable Approximation Scheme}

For any practical calculation one has to truncate the scheme. In the
weak-coupling limit, the perturbative expansion may be restricted to a certain
order. Then no particular problems are encountered as far as conservation laws
are concerned, since they are fulfilled order by order in perturbation theory.
On the other hand, such perturbative expansion may not be adequate, as, for
example, in the strong coupling limit, where resummation concepts have to be
applied. Such schemes sum up certain sub-series of diagrams to any order.
Furthermore, with the aim to solve dynamical equations of motion, such as
transport equations, one automatically resums all terms in the equations of
motion to any order. For such resumming schemes, the situation with
conservation laws is not quite as clear.

We consider the so-called $\Phi$-derivable approximations, first introduced by
Baym \cite{Baym} within the imaginary time formulation.  A $\Phi$-derivable
approximation is constructed by confining the infinite set of diagrams for
$\Phi$ to either only a few of them or some sub-series of them. Note that
$\Phi$ itself is constructed in terms of ``full'' Green's functions, where
``full'' now takes the sense of solving self-consistently the Dyson's equation
with the driving terms derived from this $\Phi$ through relation
(\ref{varphi}). It means that even restricting ourselves to a single diagram
in $\Phi$, in fact, we deal with a whole sub-series of diagrams in terms of
free Green's functions, and ``full'' takes the sense of the sum of this whole
sub-series.  Thus, a $\Phi$-derivable approximation offers a natural way of
introducing closed, i.e. consistent approximation schemes based on summation
of diagrammatic sub-series. In order to preserve the original symmetry of the
exact $\Phi$, we postulate that the set of diagrams defining the
$\Phi$-derivable approximation complies with all such symmetries. As a
consequence, approximate forms of $\Phi^{\scr{(appr.)}}$ define {\em
  effective} theories, where $\Phi^{\scr{(appr.)}}$ serves as a generating
functional for approximate self-energies $\Sa^{\scr{(appr.)}}(x,y)$
%
\begin{equation}\label{varphdl1/appr}
-\ii
\Se^{\scr{(appr.)}}(x,y)= \mp  
\frac{\delta\ii \Phi^{\scr{(appr.)}}}{\delta \ii\Gr^{\scr{(appr.)}}(y,x)}
\end{equation}
%
which then enter as driving terms for the Dyson's equations (\ref{Dyson}). The
propagators, solving this set of Dyson's equations, are still called ``full''
in the sense of the $\Phi^{\scr{(appr.)}}$-derivable scheme.  Below, we omit
the superscript ``appr.''.

\section{Generalized Kinetic Equation}\label{sect-Kin-EqT}  
\subsection{Gradient Expansion Scheme}

For slightly inhomogeneous and slowly evolving systems, the degrees of freedom
can be subdivided into rapid and slow ones. Any kinetic approximation is
essentially based on this assumption.  Then for any two-point function
$F(x,y)$, one separates the variable $\xi =(t_1-t_2, \vec{r_1}-\vec{r_2})$,
which relates to rapid and short-ranged microscopic processes, and the
variable $X= \frac{1}{2}(t_1+t_2,\vec{r_1}+\vec{r_2})$, which refers to slow
and long-ranged collective motions. The Wigner transformation, i.e.  the
Fourier transformation in four-space difference $\xi=x-y$ to four-momentum $p$
leads to the corresponding Wigner densities in four-phase-space. Since the
Wigner transformation is defined for physical space-time coordinates rather
than for contour coordinates one has to decompose the contour integrations
into its two branches, the time-ordered $\{-\}$ branch and the anti-time
ordered $\{+\}$ branch. This is explained in detail in Appendix \ref{Contour}.
Two-point functions then become matrices of the contour decomposed $\{-+\}$
components with physical space-time arguments. Thus
%
\begin{equation}
\label{W-transf} 
F^{ij}(X;p)=\int \di \xi e^{\ii p\xi}
F^{ij}\left(X+\xi/2,X-\xi/2\right),
\quad\quad\quad i,j\in\{-+\}
\end{equation}
%
leads to a four-phase-space representation of two-point functions.  The
retarded or advanced relations between two-point functions formulated on
real-time contour and those in matrix notation, $F^{ij}$, presented in
Appendix \ref{Contour}, are thereby preserved.

The Wigner transformation of Dyson's equation (\ref{Dyson}) is straight
forward.  Taking the difference and half-sum of Dyson's equation (\ref{Dyson})
and the corresponding adjoint equation and doing the Wigner transformation we
arrive at equations
%
\begin{eqnarray}
\label{Dyson-}
\ii \vu_{\mu}  \partial^\mu_X \Gr^{ij}(X,p)
&=& 
\int \di \xi e^{\ii p\xi} \oint \di z 
\left( \Se(x^i,z)\Gr(z,y^j) - \Gr(x^i,z)\Se(z,y^j) \right), 
\\
\label{Dyson+}
\widehat{Q}_X \Gr^{ij}(X,p)
&=&
\sigma^{ij} +
\frac{1}{2} \int \di \xi e^{\ii p\xi} \oint \di z 
\left( \Se(x^i,z)\Gr(z,y^j) + \Gr(x^i,z)\Se(z,y^j) \right), 
\end{eqnarray}
%
where $\sigma^{ij}$ accounts for the integration sense on the two contour
branches, cf.  eqs (\ref{sig}), 
(\ref{Fij}). For nonrelativistic kinematics $\vu^{\mu}=(1,
\vec{p} /m)$, and $\widehat{Q}_X = p_0-{\vec p}^2/2m- \partial_{\vecs
  X}^2/8m$.  In this matrix notation, two of equations (\ref{Dyson-}) and
(\ref{Dyson+}), involving $\Gr^{-+}$ and $\Gr^{+-}$ on the left-hand side, are
known as Kadanoff-Baym equations in Wigner representation \cite{Kad62}.
Particular combinations of these equations lead to the retarded and advanced
equations which completely decouple and involve only integrations over
physical times rather than contour times.

Standard transport descriptions usually involve two further approximation
steps: (i) the gradient expansion for the slow degrees of freedom, as well as
(ii) the quasiparticle approximation for rapid ones. We intend to avoid the
latter approximation and will solely deal with the gradient approximation for
slow collective motions by performing the gradient expansion of eqs
(\ref{Dyson-}) and (\ref{Dyson+}). This step preserves all the invariances of
the $\Phi$ functional in a $\Phi$-derivable approximation.

Technically, the gradient expansion is performed as follows.  The Wigner
transformation of a convolution of two two-point functions is given by
%
\begin{equation}\label{CK}
\int \di \xi e^{\ii p \xi}
\left( 
\int \di z f(x,z) \varphi(z,y)
\right) = 
\left(
\exp\left[\frac{\ii\hbar}{2}\left(\partial_p \partial_{X'}-
\partial_X \partial_{p'}\right)\right] f(X, p) \varphi (X',p')
\right)_{p'=p,X'=X}.
\end{equation}
%
Here, we have temporally restored $\hbar$ for the sake of physical lucidity. 
Expanding this expression in powers of $\hbar\partial_X \cdot
\partial_p$ to the first order, we obtain
%
\begin{equation}
\label{g-rule2}
\int \di \xi e^{\ii p \xi}
\left( \int \di z f(x,z) \varphi(z,y) \right)  
\simeq 
f(X,p) \varphi(X,p) +
\frac{\ii\hbar}{2} 
\Pbr{f(X,p), \varphi(X,p)}, 
\end{equation}
%
where  
%
\begin{equation}
\label{[]} 
\Pbr{f(X,p) , \varphi(X,p)} = 
\frac{\partial f}{\partial p^{\mu}}  
\frac{\partial \varphi}{\partial X_{\mu}} 
- 
\frac{\partial f}{\partial X^{\mu}}  
\frac{\partial \varphi}{\partial p_{\mu}} 
\end{equation}
%
is the classical Poisson bracket in covariant notation.  We would like to
stress that the smallness of the $\hbar\partial_X \cdot\partial_p$ comes
solely from the smallness of space--time gradients $\partial_X$, while
momentum derivatives $\partial_p$ are not assumed to be small!  They must be
not too large only in order to preserve the smallness of the total parameter
$\hbar\partial_X \cdot\partial_p$. This point is sometimes incorrectly
treated in the literature.

\subsection{Kinetic Equation in Physical Notation}\label{KEiPN}

It is helpful to avoid all the imaginary factors inherent in the standard
Green's function formulation and introduce two quantities which are real and,
in the quasi-homogeneous limit, positive, cf. subsect. \ref{Positive
Def}, 
and therefore have a straightforward
physical interpretation, much like for the Boltzmann equation.  We define
%
\begin{eqnarray}
\label{F}
\Fd (X,p) &=& \A (X,p) \fd (X,p)
 =  (\mp )\ii \Gr^{-+} (X,p) , \nonumber\\
\Fdt (X,p) &=& \A (X,p) [1 \mp \fd (X,p)] = \ii \Gr^{+-} (X,p) , 
\end{eqnarray}
%
for the generalized Wigner functions $\F$ and $\Ft$ and the corresponding {\em
  four}-phase-space distribution functions $\fd(X,p)$ and Fermi/Bose factors
$[1 \mp \fd (X,p)]$. Here
%
\begin{eqnarray}
\label{A}
 A (X,p) \equiv -2\Im \Gr^R (X,p) = \Fdt \pm \Fd =
\ii \left(\Ga^{+-}-\Ga^{-+}\right)
\end{eqnarray}
is the spectral function, where $\Gr^R$ is the retarded propagator, cf. eq.
(\ref{Fretarded}) of Appendix \ref{Contour}. According to relations
(\ref{Fretarded}) and (\ref{ComplexConjugate}) between Green's functions
$\Gr^{ij}$, {\em only two real functions of all these $\Gr^{ij}$ are required
  for a complete description of the system's evolution}.

The reduced gain and loss rates of the collision integral are defined as
%
\def\ga{\gamma}
\begin{eqnarray}
\label{gain}
\Ldt (X,p) &=&\Gamma (X,p)\ga(X,p)=   (\mp )\ii \Se^{-+} (X,p),\\
\Ld (X,p)  &=&\Gamma (X,p)[1\mp\ga(X,p)]=  \ii \Se^{+-} (X,p)  
\end{eqnarray}
%
with the damping width
%
\begin{eqnarray}
\label{G-def}
\Gamma (X,p)&\equiv& -2\Im \Se^R (X,p) = \Ld (X,p)\pm\Ldt (X,p), 
\end{eqnarray}
%
where $\Se^R$ is the retarded self-energy, cf. eq. (\ref{Fretarded}). 
The opposite combination
%
\begin{eqnarray}
\label{Fluc-def}
I (X,p) = [2\ga\mp1]\Gamma=\Ldt (X,p)\mp\Ld (X,p), 
\end{eqnarray}
%
is related to fluctuations. The dimensionless quantity $\ga$ is introduced for further
convenience.
 
In terms of the new notation (\ref{F})--(\ref{G-def}) and within the
first-order gradient approximation, the equations (\ref{Dyson-}) for $\Fd$ and
$\Fdt$ take the form
%
\begin{eqnarray}
\label{keqk1}
\Do 
\Fd (X,p) - 
\Pbr{\Ldt , \Re\Gr^R}
&=& C (X,p) , 
\\\label{keqkt1}
\Do 
\Fdt (X,p) - 
\Pbr{\Ld , \Re\Gr^R}
&=&\mp C (X,p) 
\end{eqnarray}
%
which we denote as the transport equations.  Here the differential drift
operator is defined as
%
\begin{eqnarray}\label{Drift-O}
\Do = 
\left(
\vu_{\mu} - 
\frac{\partial \Re\Sa^R}{\partial p^{\mu}} 
\right) 
\partial^{\mu}_X + 
\frac{\partial \Re\Sa^R}{\partial X^{\mu}}  
\frac{\partial }{\partial p_{\mu}}, \quad
v^\mu=(1,\vec{p}/m) .
\end{eqnarray}
Furthermore  
%
\begin{eqnarray}
\label{Coll(kin)}
C (X,p) =
\Ldt (X,p) \Ft (X,p) 
- \Ld (X,p) \F (X,p)=A\Gamma[\ga-f] 
\end{eqnarray}
%
is the collision term with the dimensionless functions $f$ and $\ga$ defined
in (\ref{F}) and (\ref{gain}), while $\Pbr{\Ldt , \Re\Gr^R}$ and $\Pbr{\Ld ,
  \Re\Gr^R}$ are the fluctuation terms.  Within the same approximation level
eq.  (\ref{Dyson+}) provides us with two alternative equations for $\Fd$ and
$\Fdt$
%
\begin{eqnarray}
\label{mseq(k)1}
M\Fd - \Re\Ga^R\Ldt
&=&\frac{1}{4}\left(\Pbr{\Gm,\Fd} - \Pbr{\Ldt,\A}\right),
\\\label{mseqt(k)1}
M\Fdt - \Re\Ga^R\Ld
&=&\frac{1}{4}\left(\Pbr{\Gm,\Fdt} - \Pbr{\Ld,\A}\right)
\end{eqnarray}
%
with the  ''mass'' function 
%
\begin{eqnarray}\label{meqx}\label{M}
M(X,p)=p_0 -\frac{1}{2m}\vec{p}^2 -\Re\Se^R (X,p), 
\end{eqnarray}
which relates to the drift operator via $\Do f=\Pbr{M,f}$ for any
four-phase-space function $f$. Eqs (\ref{mseq(k)1}), (\ref{mseqt(k)1}) 
are reported as the mass
shell equations. Appropriate combinations of the two sets of equations
(\ref{keqk1})--(\ref{keqkt1}) and (\ref{mseq(k)1})--(\ref{mseqt(k)1}) provide
us with the retarded equations
%
\begin{eqnarray}
\label{keqX}
\Do \Ga^R (X,p) + \frac{\ii}{2} \Pbr{\Gm , \Ga^R} &=&0, \\ 
\label{meqX}
\left( M (X,p)+\frac{\ii}{2} \Gm (X,p) \right)\Ga^R (X,p) &=& 1 . 
\end{eqnarray}
%
This subset (\ref{keqX})--(\ref{meqX}) is
solved by \cite{Bot90}
%
\begin{eqnarray}
\label{Asol}
\Gr^R=\frac{1}{M(X,p)+\ii\Gamma(X,p)/2}\Rightarrow
\left\{\begin{array}{rcl}
A (X,p) &=&\displaystyle
\frac{\Gamma (X,p)}{M^2 (X,p) + \Gamma^2 (X,p) /4},\\[4mm]
\Re\Gr^R (X,p) &=& \displaystyle 
\frac{M (X,p)}{M^2 (X,p) + \Gamma^2 (X,p) /4}.
\end{array}\right. 
\end{eqnarray}
%
The spectral function satisfies the sum--rule
\begin{eqnarray}
\label{A-sumf} 
\int_{-\infty}^{\infty} \frac{\di p_0}{2\pi} 
A(X,p) &=& 1 , 
\end{eqnarray}
%
which follows from the canonical equal-time (anti) commutation relations
for (fermionic) bosonic field operators.

With the solution (\ref{Asol}) for $\Gr^R$ equations (\ref{keqk1}) and
(\ref{mseq(k)1}) become identical to (\ref{keqkt1}) and (\ref{mseqt(k)1}),
respectively. However, equations (\ref{keqk1}) and (\ref{mseq(k)1}) are
nonidentical, while they were identical before the gradient expansion.
We will now discuss under which conditions they are equivalent. First, this
takes place in the standard quasiparticle limit when one puts
$\Gamma\rightarrow 0$ in the Green's functions.  This limit will be discussed
in the next subsect. The equivalence is also given in the case we are
interested in here, where $\Gamma $ is of finite value, however, one needs
$|f-\ga|\ll 1$ which does not necessarily require the system to be close to
local equilibrium.  In both cases the gradient approximation implies that the
macroscopic time scale $\tau_{macro}$, characterizing kinetic processes, is
much larger than the microscopic time scale $\tau_{micro}$, relating to rapid
microscopic processes.

Using the limit $|f(X,p)-\ga(X,p)|\ll 1$ and writing all $\Ldt=\ga\Gamma$
terms as $(f+(\ga-f))\Gamma$, the transport equation (\ref{keqk1}) and
the 
mass-shell equation  (\ref{mseq(k)1}) take the forms
%
\begin{eqnarray}\label{keqk2}
\frac{A^2}{2}\left(\Gamma\Pbr{M,f}-M\Pbr{\Gamma,f}\right)
  &=&\Gamma A(\ga-f)-r_{\mathrm{kin}}, \\
\frac{A\Re \Ga^R}{2}\left(\Gamma\Pbr{M,f}-M\Pbr{\Gamma,f}\right)
  &=& M A(\ga-f)-r_{\mathrm{mass-eq}},\label{mseq(k)2}
\end{eqnarray}
%
where the remaining terms $r_{\mathrm{kin}}$ and $r_{\mathrm{mass-shell}}$ are
of order $(\ga-f)$ times gradient terms, i.e. of second order in the gradient
expansion\footnote{These terms have the explicit form
  \begin{eqnarray}
    r_{\mathrm{kin}}
    =
-\Pbr{\Gamma (\ga-f),\Re \Ga^R },\,\,\,
    r_{\mathrm{mass-eq}}
    = \frac{1}{4}\Pbr{\Gamma (\ga-f), A }.
 \end{eqnarray}}. 
From eq. (\ref{Asol}) one sees that 
$\Gamma\Re\Gr^R=MA$.  
Using this we observe that the mass-shell equation
loses its original sense, since to leading order relation (\ref{mseq(k)2}) is
equivalent to the kinetic equation (\ref{keqk2}). However, the still remaining
difference in the second-order terms is inconvenient from the practical point
of view. Following Botermans and Malfliet \cite{Bot90} who first 
suggested to drop the
above $r_{kin}$ term in the kinetic equation we put $r_{kin}=r_{mass}=0$, since
then both equations indeed become identical. This step amounts to replace the
$\Ldt$ and $\Ld$ terms by $f\Gamma$ and $(1\mp f)\Gamma$ in {\em all Poisson
  brackets}. The so obtained {\em generalized kinetic equations} for $\Fd$ and
$\Fdt$ then read
%
\begin{eqnarray}
\label{keqk}
\Do 
\Fd (X,p) - 
\Pbr{\Gm\frac{\Fd}{\A},\Re\Gr^R} &=& C (X,p), \\ 
\label{keqkt}
\Do 
\Fdt (X,p) - 
\Pbr{\Gm\frac{\Fdt}{\A},\Re\Gr^R} &=& \mp C (X,p).  
\end{eqnarray}
%
Although these two equations are identical to each other, we nevertheless have
presented both of them for the sake of further convenience. With the help of
the retarded relations (\ref{Asol}) both equations can be converted to one
generalized transport equation for the phase-space occupation functions $f$
%
\begin{eqnarray}
\label{keqk3}
\frac{A^2}{2}\left(\Gamma\Do f-M\Pbr{\Gamma,f}\right)
  &=& C .  
\end{eqnarray}  
%
To get eq. (\ref{keqk3}) we used that the additional Poisson-bracket
fluctuation
term becomes
%
\begin{eqnarray}
\label{backflow}  
\Pbr{\Gm f,\Re\Gr^R}=\frac{M^2-\Gamma^2/4}{(M^2+\Gamma^2/4)^2}\;
   \Do\left(\Gm f\right)
   +\frac{M\Gamma^2 /2}{(M^2+\Gamma^2/4)^2}\Pbr{\Gm,f}.
\end{eqnarray}
%

We now provide a physical interpretation of various terms in the generalized
kinetic equation (\ref{keqk}) or (\ref{keqk3}).  This physical interpretation
relies on the similarity of most of the terms to conventional kinetic
equations, for example, such as the Landau equation for Fermi liquids (see,
e.g., refs \cite{Baym91,Lif81}) and ref.  \cite{BCh} for the relativistic
version), proposed by Landau on the basis of an intuitive quasiparticle
picture \cite{Land56}.

For this purpose it is advantage to convert the drift operator (\ref{Drift-O})
into a space and time separated form 
%
\begin{equation}
\label{Do-qp-1}
\Do=
\frac{1}{Z} \left(\partial_t   + 
 \vec{v}_g 
\partial_{\bf X} \right)  + 
\partial_t \Re\Sa^R \cdot {\partial_{p_0}} - 
\partial_{\bf X} \Re\Sa^R \cdot {\partial_{\vec p}}\;,
\end{equation}
%
where
%
\begin{eqnarray}
\label{3-vel}\label{Z-norm}  
\vec{v}_g (X,p) = Z
\left(\vec\vu + \frac{\partial \Re\Se^R}{\partial \vec p}\right)
\quad\mbox{with}\quad
Z =\left(\vu_0   - \frac{\partial \Re\Se^R}{\partial p_0}\right)^{-1}
\end{eqnarray}
%
takes the meaning of the group 3-velocity in the quasiparticle approximation,
while $Z$ is the standard renormalization factor.

Thus, the drift term $\Do \Fd$ on the l.h.s. of eq. (\ref{keqk}) is the usual
kinetic drift term including the corrections from the self-consistent field
$\Re\Se^R$ into the convective transfer of real and also virtual particles.
In the collisionless case \mbox{$\Do \Fd=0$} (Vlasov
equation), the quasi-linear first-order differential operator $\Do$ defines
characteristic curves. They are the standard classical paths.  The
four-phase-space probability $\Fd(X,p)$ is conserved along these paths. The
formulation in terms of a Poisson bracket in four dimensions implies a
generalized Liouville theorem. In the collisional case, both the collision, $C$,
and fluctuation terms (\ref{backflow}) change the phase-space probabilities of
the ``generalized'' particles during their propagation along the
``generalized'' classical paths given by $\Do$. We use the term
``generalized'' in order to emphasize that particles are no longer bound to
their mass-shell, $M=0$, during propagation due to the collision term, i.e.
due to decay, creation or scattering processes.

The r.h.s. of eq. (\ref{keqk}) specifies the collision term $C$ in terms of
gain and loss terms, which also can account for multi-particle processes.
Since $\Fd$ includes a factor $A$, $C$ further deviates from the standard
Boltzmann-type form in as much that it is multiplied by the spectral function
$A$, which accounts for the finite width of the particles.  The
Poisson-bracket term (\ref{backflow}) is special.  It contains genuine
contributions from the finite mass width of the particles and describes the
response of the surrounding matter due to fluctuations. This can be seen from
the conservation laws discussed below. In particular the first term in
(\ref{backflow}) gives rise to a back-flow component of the surrounding
matter. It restores the Noether currents as the conserved ones rather than the
intuitively expected sum of convective currents arising from the convective
$\Do\F$ terms in (\ref{keqk}). In eq. (\ref{keqk3}) one further realizes that
the function $\frac{1}{2}A^2\Gamma$ in front of the drift term is more sharply
peaked than the original spectral function $A$. Both
$\frac{1}{2}A^2\Gamma$ and  $A$ are reduced to the same
$\delta$-function in the quasiparticle limit.
The $M\Pbr{\Gamma,f}$ term,
which gives no contribution in the quasiparticle limit due to the factor $M$,
represents a specific off-mass-shell response, cf.  ref. \cite{LipS}.

So far the gradient approximation has been applied to the space-time foldings
occurring between the self-energies and the propagators appearing in the
collision term. They give rise to the gradient terms on the l.h.s. of the
kinetic equations. This is sufficient as long as the self-energies are
calculated without further approximation. Commonly one likes to obtain also
$\Ldt$ and $\Ld$ in a kind of local approximation evaluated with all Green's
functions taken at the same space-time point $X$. Then for self-energy diagrams
with more than two points, also nonlocal gradient corrections within these
diagrams arise, which have to be accounted for in a consistent gradient
approximation scheme. The latter fact gives rise to memory effects, which will
be discussed in detail in subsect. \ref{LocalCol}, below.

In all derivations above the $\Ldt$ and $\Ld$ terms in all Poisson brackets
have consistently been replaced by $f\Gamma$ and $(1\mp f)\Gamma$,
respectively. In fact, this procedure is optional. However, in doing so we
gained few advantages. First, all our generalized kinetic equations,
(\ref{keqk2}), and mass-shell equations, (\ref{mseq(k)2}) are {\em exactly}
equivalent to each other, as they were in exact theory before the gradient
expansion.  Second, the so obtained generalized kinetic equation (\ref{keqk3})
has particular feature with respect to the definition of a nonequilibrium
entropy flow in connection with the formulation of an {\em exact} H-theorem in
certain cases (sect.  \ref{H-theorem}).  If we refrained from these
substitutions, these both features would become approximate with the accuracy
up to second-order gradients. While this accuracy still complies with the
gradient approximation level, it makes our first-order approximation less
elegant.

\subsection{Quasiparticle Limit}\label{QP}

The quasiparticle picture is recovered in the limit of small (while nonzero)
width $\Gamma$.  Small means that $\Gamma$ should be less than all typical
scales in the problem. This allows to put $\Gamma$ zero in all Green's
functions which are, thereby, called quasiparticle Green's functions.
nonzero means that $\Gamma$ is kept finite in the collision term being now
calculated with the help of these quasiparticle Green's functions.
 
Thus, in this limit the spectral function (\ref{Asol}) takes the form
%
\begin{equation}
\label{dyn-qp-lim} 
A^{\scr{qp}} =2\pi\; \sgn \Gamma \; \delta(M) 
\end{equation}
%
with $M$ given by eq. (\ref{M}). As the right hand side of the mass shell
equation (\ref{mseq(k)1}) contains one $\Gamma$ factor more than the left
hand side, it reduces to $MA(f-\ga)=0$ requiring $M$ to vanish within the
range of the spectral function. This just complies with the on-shell
$\delta$-function in eq.  (\ref{dyn-qp-lim}) which with $M=0$ provides the
dispersion equation for the quasiparticle energies $\varepsilon (X,\vec p)$
%
\begin{equation}
\label{qp-disp-non}
\varepsilon (X,\vec p)=\frac{1}{2m}{\vec p}^2
+
\Re\;\Se^R\left(X,\varepsilon (X,\vec p),\vec p\right).
\end{equation}
%
This dispersion equation may have several solutions (branches). For brevity we
drop a possible branch index. In the quasiparticle limit the generalized
Wigner functions and the Fermi/Bose factors (\ref{F}) are presented in the
form
%
\begin{eqnarray}
\label{F-q.p.}
\F^{\scr{qp}} (X,p)&=& 2\pi\; Z^{\scr{qp}} (X,\vec p )\; 
\delta \left( p_0 - \varepsilon (X,\vec p)\right) 
f^{\scr{qp}}(X, \vec p ), \\
\label{Ft-q.p.}
\Ft^{\scr{qp}} (X,p)&=& 2\pi\; Z^{\scr{qp}} (X,\vec p )\; 
\delta \left( p_0 - \varepsilon (X,\vec p)\right) 
\left[1\mp f^{\scr{qp}}(X, \vec p )\right], 
\end{eqnarray}
%
with the quasiparticle distribution functions 
%
\begin{equation}
\label{qp-distr}
f^{\scr{qp}}(X, \vec p ) =  
f \left(X,\varepsilon (X,\vec p),\vec p \right),
\end{equation}
%
%
which depend on the 3-momentum only rather than on the 4-momentum,
while
%
\begin{equation}
\label{Z-q.p.}
Z^{\scr{qp}} (X,\vec p ) = 
Z \left(X,\varepsilon (X,\vec p),\vec p \right) 
\end{equation}
%
denotes the quasiparticle normalization factor (cf. eq. (\ref{Z-norm})).  In
practice, $Z^{\scr{qp}}$ is always positive for truly nonrelativistic systems
\cite{KSK94}.  We have replaced $p_0$ by $\varepsilon (X,\vec p)$ in all
relations above due to the on-shell $\delta$-function in eq.
(\ref{dyn-qp-lim}).

Integration of eq. (\ref{keqk}) over $p_0$ in a narrow region of width $\Gm$
around the branch position $\varepsilon (X,\vec p)$, where one can identify
$A=A^{\scr{qp}}$, leads to the standard quasiparticle transport equation
%
\begin{equation}
\label{q.p.-kin}
\partial_t f^{\scr{qp}} (X,\vec p)+ 
\frac{\partial \varepsilon (X,\vec p)}{\partial {\vec p}}
\partial_{\bf X} f^{\scr{qp}}(X,\vec p) - 
\partial_{\bf X} \varepsilon (X,\vec p) \;
\frac{\partial f^{\scr{qp}}(X,\vec p)}{\partial {\vec p}} 
= C^{\scr{qp}}(X,\vec p). 
\end{equation}
%
Here $C^{\scr{qp}}(X,\vec p)$ is the corresponding collision integral now
expressed in terms of quasiparticle distributions. The fluctuation term
$\Pbr{\Gm f,\Re\Gr^R}$ integrated over $p_0$ becomes of second order in $\Gm$
and thus drops out of this quasiparticle equation.

This quasiparticle transport equation is consistent as long as its evolution
does not lead beyond its region of validity, i.e. into region where $\Gm$
becomes comparable to typical energy scales in the problem. For instance, the
consistency of the Landau Fermi-liquid theory is preserved, provided only
low-excited Fermi systems are considered. Then the Pauli-blocking prevents
particles to scatter to states laying far away from the Fermi surface, i.e. to
a region where the quasiparticle approximation would be violated.

In above approach neither any uncontrolled approximations nor any specific
ansatz has been introduced.  The quasiparticle limit of the generalized
kinetic equation directly follows from our consideration and encounters no
problems for the eliminating the extra Poisson bracket term.  Here a caution
should be expressed. The quasiparticle form of the spectral function of eq.
(\ref{dyn-qp-lim}) gives
%
\begin{equation}
\label{qp-sumr}
\int_{-\infty}^{\infty}\frac{\di p_0}{2\pi} \A^{\scr{qp}} (X,p) =
Z^{\scr{qp}} (X,\vec p ) 
\end{equation}
%
for the sum-rule rather than unity. The $Z^{\scr{qp}}$ factor appears here,
because in (\ref{qp-sumr}) we artificially extended the quasiparticle form of
$A$ to all energies and did not account for background terms outside the
quasiparticle region. However, in this form the sum rule is often taken as the
canonical one for quasiparticles, cf. ref. \cite{Neg88}, in order to construct
an effective theory only in terms of quasiparticles. This recipe works, if the
quasiparticle energy region is indeed well separated from the region, where
particles have a rather large width.

\subsection{Conservations of Charge and Energy--Momentum}
\label{Conservation-L}

Transport equation (\ref{keqk}) weighted either with the charge $e$ or with
4-momentum $p^\nu$, integrated over momentum and summed over internal degrees
of freedom like spin ($\Tr$) gives rise to the charge or energy--momentum
conservation laws, respectively, with the Noether 4-current and Noether
energy--momentum tensor defined by the following expressions
%
\begin{eqnarray}
\label{c-new-currentk} 
j^{\mu} (X) 
&=& e \mbox{Tr} \int \dpi{p}
\vu^{\mu} 
\Fd (X,p), \\
\label{E-M-new-tensork}
\Theta^{\mu\nu}(X)
&=&
\mbox{Tr} \int \dpi{p} 
\vu^{\mu} p^{\nu} \Fd (X,p)
+ g^{\mu\nu}\left(
{\cal E}^{\scr{int}}(X)-{\cal E}^{\scr{pot}}(X)
\right).  
\end{eqnarray}
%
Here 
%
\begin{eqnarray}
\label{eps-int} 
{\cal E}^{\scr{int}}(X)=\left<-\Lint(X)\right>
=\left.\frac{\delta\Phi}{\delta\lambda(x)}\right|_{\lambda=1}
\end{eqnarray}
%
is interaction energy density, which in terms of $\Phi$ is given by a
functional variation with respect to a space-time dependent coupling strength
of interaction part of the Lagrangian density 
$\Lint\rightarrow\lambda(x)\Lint$, cf. ref. \cite{IKV}.  The potential
energy density ${\cal E}^{\scr{pot}}$ takes the form
%
\begin{eqnarray}
\label{eps-potk}
{\cal E}^{\scr{pot}}
= 
\mbox{Tr}
\int\dpi{p} \left[
\Re\Sa^R \Fd
+ \Re\Ga^R \frac{\Gm}{\A}\F
\right].  
\end{eqnarray}
%
Whereas the first term in squared brackets complies with quasiparticle
expectations, namely mean potential times density, the second term displays
the role of fluctuations in the potential energy density.

For specific interactions with the same number $\alpha$ of field operators
attached to any vertex of $\Lint$, one simply deduces
\begin{eqnarray}\label{int-spec}
{\cal E}^{\scr{int}}(X,p)=\frac{2}{\alpha}{\cal E}^{\scr{pot}}(X,p).
\end{eqnarray}
Please also notice from (\ref{E-M-new-tensork}) that the special combination 
\begin{eqnarray}
\label{Energy-Pressure}
\Theta^{00}(X)+
\frac{1}{3}\sum_{i=1}^3\Theta^{ii}(X)=
\mbox{Tr} \int \dpi{p} 
A (X,p)f(X,p)\left[ p_0 - \frac{2}{3}\epsilon_p^{0} \right]
\end{eqnarray}
depends on the specific form of the interaction only via the spectral
function.  As we will see below, this combination simply relates to the
entropy density in local thermodynamic equilibrium.

The conservation laws only hold, if all the self-energies are
$\Phi$-derivable. In ref. \cite{IKV}, it has been shown that this implies the
following consistency relations obtained after Wigner transformation and
first-order gradient expansion, (a) for the conserved current
%
\begin{equation}
\label{invarJk}
\ii \mbox{Tr} \int \dpi{p} 
\left[
\Pbr{\Re\Sa^R,
\Fd} 
- 
\Pbr{\Re\Ga^R,\frac{\Gm}{\A}
\F}  
+ C
\right]
=0, 
\end{equation}
%
and (b) for the energy-momentum tensor
%
\begin{eqnarray}
\label{epsilon-invk}
\partial^{\nu}
\left(
{\cal E}^{\scr{int}} - {\cal E}^{\scr{pot}}
\right)
= \mbox{Tr}
\int 
\frac{p^\nu \di^4 p}{(2\pi )^4}
\left[
\Pbr{\Re\Sa^R,
\Fd} 
- 
\Pbr{\Re\Ga^R,\frac{\Gm}{\A}
\F}  
+C
\right] .
\end{eqnarray}
%
The contributions from the Markovian collision term $C$ drop out in both
cases, cf. eq. (\ref{Multi-rate}) below.  The first term in each of the two
relations refers to the change from the free velocity $\vec v$ to the group
velocity $\vec v_g$, cf. eq. (\ref{3-vel}), in the medium. It can therefore be
associated with a corresponding {\em drag--flow} contribution of the
surrounding matter to the current or energy--momentum flow. The second
(fluctuation) term compensates the former contribution and can therefore be
associated with a {\em back--flow} contribution, which restores the Noether
expressions (\ref{c-new-currentk}) and (\ref{E-M-new-tensork}) to be indeed
the conserved quantities. In this compensation we see the essential role of
the fluctuation term (\ref{backflow}) in the generalized kinetic equation.
Dropping or approximating this term would spoil the conservation laws. Indeed,
both expressions (\ref{c-new-currentk}) and (\ref{E-M-new-tensork}) comply
exactly with the generalized kinetic equation (\ref{keqk}), i.e. they are
exact integrals of the generalized kinetic equations.

Expressions (\ref{c-new-currentk}) and (\ref{E-M-new-tensork}) for 4-current
end energy--momentum tensor, respectively, as well as self-consistency
relations (\ref{invarJk}) and (\ref{epsilon-invk}) are written explicitly for
the case of nonrelativistic particles which number is conserved. This follows
from the conventional way of nonrelativistic renormalization for such
particles based on normal ordering. When the number of particles is not
conserved (e.g., for phonons) or a system of relativistic particles is
considered, one should replace $\Fd (X,p) \rightarrow \frac{1}{2}\left(\Fd
  (X,p) \mp \Fdt (X,p) \right)$ in all above formulas in order to take proper
account of zero point vibrations (e.g., of phonons) or of the vacuum
polarization in the relativistic case.  These symmetrized equations, derived
from special ($\mp$) combinations of the transport equations (\ref{keqk}) and
(\ref{keqkt}), are generally ultra-violet divergent, and hence, have to be
properly renormalized at the vacuum level.

\subsection{Positive Definiteness of Kinetic Quantities} 
\label{Positive Def}

For a semi-classical interpretation one likes to have the Wigner distributions
$\F(X,p)$ and $\Ft(X,p)$ to be positive semi-definite, hereto after just
called ''positive''. Using the operator definition for the Green's functions
(\ref{Ga1}) and integrating it over a large space-time volume, one arrives at,
e.g.,
%
\begin{eqnarray}
\label{Gr-x-int}
\int \di X \F(X,p) = 
\left<\left(\int \di y e^{\ii p y}\Ptd(y)\right) 
\left(\int \di x e^{-\ii p x}\Pt(x)\right)\right>, 
\end{eqnarray}
%
and similarly for self-energies $\Se$ expressed through the current--current
correlator. It indicates that the r.h.s. of such equality is real and
nonnegative. Thus, we get the following set of constraints
%
\begin{eqnarray}
\label{Gr{+-}>0}\hspace*{-0.7cm}
\int \di X \Ft(X,p)  \geq 0, \,\,\,
\int \di X \F(X,p)  \geq 0,\,\,\, 
\int \di X \Get(X,p)  \geq 0, \,\,\,
\int \di X \Ge(X,p)  \geq 0. 
\end{eqnarray}
%
Similar relations are obtained for the integration over four-momentum space
rather than space and time. As a result, in stationary and spatially
homogeneous systems, in particular in equilibrium systems, the quantities
$\Fd$, $\Fdt$, $\Gb$ and $\Gbt$ are real and nonnegative, i.e.
%
\begin{eqnarray}
\label{F>0st}
\Fd(p)  \geq 0, \,\,\,\, \Fdt(p)  \geq 0;  \,\,\,\,
\Gb(p)  \geq 0, \,\,\,\, \Gbt(p)  \geq 0. 
\end{eqnarray}
%
In deriving constraints (\ref{Gr{+-}>0}) and (\ref{F>0st}), we did not use the
fact that the Green's functions are solutions of the Dyson's equation.
However, we used the operator picture. Any approximation, in particular, if
formulated in the space of Green's functions, may spoil such rigorous
statements like (\ref{Gr-x-int}). Nevertheless, both the $\Phi$-derivable
scheme and the gradient approximation preserve the retarded relations
(\ref{Fretarded}) among the different contour components and the retarded and
advanced functions of any contour function, with definite values for the
imaginary parts of the corresponding retarded Wigner functions
%
\begin{eqnarray}
\label{ImGR<0}
-2\Im \Ga^R(X,p)=A(X,p)  \geq 0,\,\,\, 
-2\Im \Se^R(X,p)=\Gamma(X,p)  \geq 0,
\end{eqnarray}
%
which even hold locally.  In particular, solution (\ref{meqX}) for the
retarded Green's function shows that all retarded relations hold locally: the
momentum part is the same as that in the homogeneous case with the space-time
coordinate $X$ as a parameter. Under the condition $|f-\gamma|\ll 1$, cf. the
discussion around eq. (\ref{keqk2}), one finds that
%
\begin{eqnarray}\label{FeqR}
\Ld(X,p)\approx\fd\Gamma(X,p)>0,\quad\quad\Ldt(X,p)
\approx(1\mp\fd)\Gamma(X,p)>0
\end{eqnarray}
%
as long as the Wigner densities $f$ and $1\mp f$ are positive. As the gradient
approximation is a quasi-homogeneous approximation, one may therefore expect
the positivity of $\Ldt$ and $\Ld$ to be preserved even in the self-consistent
treatment discussed here. Diagrammatic rules may also corroborate this, since
diagrams for $\Ldt$ and $\Ld$ are calculated like in the homogeneous case.

We now like to show that, if $\Ldt$ and $\Ld$ are positive, also under minor
restrictions the kinetic equations (\ref{keqk}), (\ref{keqkt}) preserve the
positivity of $\Fd$ and $\Fdt$, once one has started from positive $\Fd$ and
$\Fdt$ initially. For this purpose we turn to the generalized kinetic
equation (\ref{keqk}) for the phase-space distribution $\Fd(X,p)$ which is of
integro-quasi-linear first-order partial differential type in 8 dimensions. We
present it in the form
%
\begin{eqnarray}\label{quasi-linear}
\Do \Fd= A\Ldt-\Gamma\Fd+\Pbr{\Gm\frac{\Fd}{\A},\Re\Gr^R},
\end{eqnarray}
%
with the coefficient $Z$ (cf. eq. (\ref{Z-norm})) in front of the time
derivative being positive. The condition $Z>0$ takes place due to the
dispersion relation for the self-energy $\Se^R$ and positive definiteness of
$\Gamma(X,p)$, cf. eq. (\ref{ImGR<0}).

Given the solution, let us then discuss properties of $\Fd$ along
characteristic curves determined by the quasi-linear drift operator $\Do$ with a
curve parameter $s$ growing monotonically with time. Assuming $\Fd$ is
positive initially, the occurrence of a negative value at some later time
requires the first zero value to occur at one of the characteristic curves.
Be $s_0$ such a place, one finds
%
\begin{eqnarray}\label{quasi-linear-so}
\frac{\di}{\di s}
\left.\Fd(s)\vphantom{\int}\right|_{\mbox{\footnotesize $s=s_0$}}=
\left[ A\Ldt+ 
\Pbr{\Gm\frac{\Fd}{\A},\Re\Gr^R}\vphantom{\int}\right]_{\mbox{\footnotesize $s=s_0$}}\; .
\end{eqnarray}
%
If the r.h.s. is positive, one comes to a contradiction, as the approach from
some positive value to a zero value which then becomes negative would require
a nonpositive derivative at $s_0$.  Thus, $A\Ldt>|\Pbr{\Gm\Fd/\A,\Re\Gr^R}|$
is a sufficient (not necessary) condition that $\Fd$ preserves its positivity.
The corresponding condition $A\Ld>|\Pbr{\Gm\Fdt/\A,\Re\Gr^R}|$
provides positivity
of $\Fdt$.  According to eq. (\ref{backflow}) this puts restrictions on the
size of the occurring gradients of the gain and loss rates along the
characteristic curves, i.e. along the classical paths. This condition is met
in a wide range of cases, namely (a) in the quasi-homogeneous and quasi-static
case, i.e. close to equilibrium, where gradients are of order $C\ll A\Ldt,
A\Ld$, and (b) in the standard quasi-particle limit, where an extra small
parameter, namely the width $\Gamma$, enters the estimations.

\section{Collision Term}
\label{Collision Term}

To further discuss the transport treatment we need an explicit form of the
collision term (\ref{Coll(kin)}), which is provided from the $\Phi$ functional
in the $-+$ matrix notation via the variation rules (\ref{var-Phi-component})
as
%
\def\tp{p'}\def\tm{m'}\def\tW{\widetilde{W}} 
\def\tR{\widetilde{R}}
\begin{eqnarray}
\label{Coll-var} 
C (X,p) =&& 
\frac{\delta\ii\Phi}{\delta\Ft(X,p)}\Ft(X,p)
-\frac{\delta\ii\Phi}{\delta\F(X,p)}\F(X,p) .
\end{eqnarray}
Here we assumed $\Phi$ be transformed into the Wigner representation and all
$\mp\ii\Gr^{-+}$ and $\ii\Gr^{+-}$ to be replaced by the Wigner-densities
$\Fd$ and $\Fdt$. Thus, the structure of the collision term can be inferred
from the structure of the diagrams contributing to the functional $\Phi$. To
this end, in close analogy to the consideration of ref.  \cite{Knoll95}, we
discuss various decompositions of the $\Phi$-functional, from which the in-
and out-rates are derived. This consideration is based on the 
standard real-time diagrammatic rules, where the contour 
integrations are decomposed into two branches with $-$ and $+$ 
vertices for the time- and anti-time-ordered branches, cf. 
Appendix \ref{Contour}. For the sake of physical transparency, 
we confine our treatment to the {\em local} case, where in 
Wigner representation all the Green's functions are taken at the 
same space-time coordinate $X$ and all nonlocalities, i.e. 
derivative corrections, are disregarded.  Derivative corrections 
give rise to memory effects in the collision term, which will be 
analyzed separately for the specific case of triangle diagram 
(see subsect.  \ref{Memory-Effects}).

\subsection{Diagrammatic Decomposition into Physical
Sub-Processes }\label{Decomposition}

Consider a given closed diagram of $\Phi$, at this level specified by a
certain number $n_{\lambda}$ of vertices and a certain contraction pattern
which links all vertices with lines of certain arrow sense for complex fields.
This fixes the topology of such a contour diagram. It leads to $2^{n_\lambda}$
different diagrams in the $-+$ notation from the summation over all $-+$ signs
attached to each vertex. Thus for any $-+$ type diagram of $\Phi$ also the
diagram, where all $+$ and $-$ vertex signs are interchanged, contributes to
$\Phi$. Furthermore, for any diagram with given line senses the diagram with
all line senses reversed also is a valid diagram of $\Phi$.  We shall exploit
these two discrete symmetry operations to further determine the properties of
$\Phi$. The simultaneous application of the interchange of all vertex signs
and the reversion of the line sense leads to the adjoint expressions, since in
the underlying operator picture it adjungates all operators and inverts the
operator ordering. The corresponding values are then complex conjugate to one
another, cf.  (\ref{ComplexConjugate}).

Using these symmetries one can convert the functional $\Phi$ into the
following general form
%
\begin{eqnarray}\label{Phi-Rmm-sum}
\ii\Phi&=&\displaystyle
     \frac{1}{2}\sum_{m,m'}\left(\ii\Phi_{m,m'}+\ii\Phi_{m',m}\right),\\
\label{Phi-Rmm}
\ii\Phi_{m,m'}&=&\displaystyle\int 
     \dpi{p_1}\cdots\dpi{p_m} \dpi{\tp_1}\cdots\dpi{\tp_{\tm}}
     \;\;\delta^4\left(\sum_{i=1}^{m} p_i - \sum_{i=1}^{\tm} p'_i \right)
     \nonumber\\ 
     &\times& R_{m,m'}(X;p_1,\dots,p_m;p'_1,\dots,p'_{m'})
     \;\Fd_1\cdots\Fd_m\Fdt'_1\cdots\Fdt'_{m'}\; ,
\end{eqnarray}
%
where in view of the local approximation the four momentum conservation has
been extracted. While $\Phi_{00}$ compiles with terms void of 
any Wigner densities, i.e. from diagrams where all vertices have 
the same sign and which do not contribute to the collision term, 
the nontrivial $\Phi_{m,m'}$ terms sum the sub-class of diagrams 
of $\Phi$ with precisely $m+m'$ Wigner densities 
$\Fd_i=F(X,p_i)$ and $\Fdt'_i=\Fdt(X,p'_i)$, respectively.  
According to eq.  (\ref{Coll-var}) each $\Phi_{m,m'}$ in 
(\ref{Phi-Rmm}) generates a multi-particle gain or loss 
contributions expressed in terms of integrals over products 
of generalized distribution functions $\Fd$, and Fermi/Bose 
factors $\Fdt$.  Every term in the sum (\ref{Phi-Rmm-sum}) has 
been duplicated, repeating each term in its line reverse form. 
The corresponding transition rates 
$R_{m,m'}(X;p_1,\dots,p_m;p'_1,\dots,p'_{m'})$ are real due to 
the adjungation symmetry. As explained below, they result from 
the product of chronological Feynman amplitudes, given by the 
sub-diagram compiled from all $-$ vertices linked by 
$\ii\Gr^{--}$ Green's functions times that of the 
anti-chronological part containing the $\ii\Gr^{++}$ functions.

The generic form (\ref{Phi-Rmm}) of $\Phi_{m,m'}$ can be illustrated
diagrammatically.  For this purpose consider a $-+$ notation diagram
contributing to $\Phi_{m,m'}$, which contains vertices of either sign.  It can
be decomposed into two pieces, denoted by a compact bra-ket notation, say
$\left(\alpha\right|$ and $\left|\beta\right)$, in such a way that each of the
two sub-pieces contains vertices of only one type of sign\footnote{To
  construct the decomposition, just deform a given mixed-vertex diagram of
  $\Phi$ in such a way that all $+$ and $-$ vertices are placed left and
  respectively right from a vertical division line and then cut along this
  line.}
%
\begin{eqnarray}\label{decomp}\unitlength6mm
\ii \Phi_{\alpha\beta}^{+-}&=&\unitlength7mm
\frac{1}{n_{\alpha\beta}}
\begin{picture}(4.5,1.5)
\put(0.,.1){\phidecomposition}
\end{picture}\vphantom{\sum_{\int}}
    =\frac{1}{n_{\alpha\beta}}\left(\alpha\left|
\Fd_1 \cdots \Fd_m \Fdt'_{1} \cdots \Fdt'_{m'}\right|\beta\right)\cr
    &=&\frac{1}{n_{\alpha\beta}}
    \int\frac{\di^{4(m+m') p_{mm'}}}{(2\pi)^{4(m+m')}}\;
    \delta^4\left(\sum_{i=1}^{m} p_i - \sum_{i=1}^{\tm} p'_i \right)\cr
    &&\hspace*{1cm}\times\;    V^*_{\alpha}(X;p_{mm'})\;
    \Fd_1 \cdots \Fd_m \Fdt'_{1} \cdots \Fdt'_{m'}\;
    V_{\beta}(X;p_{mm'}),
\end{eqnarray}
%
where $n_{\alpha\beta}$ counts the number of vertices in the closed diagram
$\left(\alpha|\beta\right)$.  The short-hand notation
$p_{mm'}=\left\{p_1,\dots,p_m;p'_{1},\dots,p'_{m'}\right\}$ summarizes the
momenta, type and internal quantum numbers of the set of ordered valences, to
be joint to the product of $m+m'$ Wigner densities\footnote{ This product of
  Wigner-densities originates from the $\mp\ii\Gr^{-+}$ and $\ii\Gr^{+-}$
  Green's functions, (cf. eq.  (\ref{F})). In closing the diagram by these
  Wigner densities extra fermion loops may appear besides the 
  ones accounted for in the amplitudes $\alpha$ and $\beta$. 
  Using the product $\Fd_1 \cdots \Fd_m \Fdt'_{1} \cdots 
  \Fdt'_{m'}$ just take care of the additional $(-1)$ factors 
  arising from this closing of the diagram.} 
$\Fd_1\cdots\Fd_m\Fdt'_1\cdots\Fdt'_{\tilde m}$ linking the two amplitudes.
The ``end-caps'' $\left(\alpha\right|$ and $\left|\beta\right)$ represent
multi-point vertex functions, in simple cases of tree type, of only one sign
for the vertices, i.e.  they are either entirely time ordered ($-$ vertices)
or entirely anti-time ordered ($+$ vertices).  Each such vertex function
%
\begin{eqnarray}\label{V}
     \left|\beta\right)&= V_{\beta}(X;p_{mm'}),\quad\quad
     \left(\alpha\right|&= V^*_{\alpha}(X;p_{mm'}),
\end{eqnarray}
%
to be determined by normal Feynman diagram rules, has $m$ $\Fd$-valences and
$m'$ $\Fdt$-valences, respectively. In (\ref{V}) we used the fact that adjoint
expressions are complex conjugate to each other, cf.
(\ref{ComplexConjugate}). Accumulating all diagrams of $\Phi$ that lead to the
same set of Wigner densities $\Fd_1\cdots\Fd_m\Fdt'_1\cdots\Fdt'_{\tilde m}$
provides us with the generic form (\ref{Phi-Rmm}) with the partial process
rates
%
\begin{equation}\label{Rmm}
R_{m,m'}(X;p_{mm'})
=
\sum_{(\alpha\beta)\in \Phi_{m,m'}}\frac{1}{n_{\alpha\beta}}
\Re\left\{
V^*_{\alpha}(X;p_{mm'})
V_{\beta}(X;p_{mm'})\right\}.
\end{equation}
%
The restriction to the real part arises, since with $(\alpha|\beta)$ also the
adjoint $(\beta|\alpha)$ diagram contributes to the subclass $\Phi_{m,m'}$.
However these rates are not necessarily positive as in perturbation
theory\footnote{In perturbation theory they are positive, since there the sum
  in (\ref{Rmm}) leads to absolute squares. In the general case with resummed
  propagators this positivity may be lost due to the restriction to diagrams of
  $\Phi$ which are globally two-particle irreducible. The latter excludes
  certain combinations of amplitude diagrams, implying that the rates of
  genuine multi-particle processes are not necessarily positive.}.  In this
point, the generalized scheme differs from the conventional Boltzmann
kinetics. Even the two rates, $R_{mm'}$ and its line sense reverse
${R_{m'm}}$, have not to be necessarily identical to each other. Still eq.
(\ref{Phi-Rmm}) represents the most general form of $\Phi$ expressed through
the Wigner densities $\Fd(X,p)$ and $\Fdt(X,p)$ in the local approximation.
It is however important to realize that in many physically relevant cases, 
e.g. those  discussed below, one indeed finds that
%
\begin{eqnarray}\label{Rmm=Rtmm}
R_{m,m'}={R}_{m',m}.
\end{eqnarray}
This property will be used as a sufficient condition for the derivation
of the H-theorem. In the following we will restrict the discussion to cases
where (\ref{Rmm=Rtmm}) is assumed.  The treatment of more general cases, in
particular in connection with the H-theorem, will be deferred to another
publication.

Since $\Phi$ is two-line irreducible, there are at least three lines
connecting $(\alpha|$ and $|\beta)$ and in many cases $(\alpha|$ and $|\beta)$ are
connected by, at least, four lines, like in the case of two-body
interactions.
In such diagrams each of the amplitudes
$\alpha$ or $\beta$ necessarily form a connected diagram for the complex field
case for binary ($m=m'=2$) and for triple scattering ($m=m'=3$) case, while in
the general case of multi-particle (more than triple) processes such
amplitudes may be disconnected.

No further symmetry can be specified at this level without additional
knowledge on possible topological and other symmetries of $\Phi$ and the
different particle species involved.  The decomposition discussed here solely
relies on a straightforward application of the contour rules for multi-point
functions.  They differ from other cutting rules like those derived by
Danielewicz \cite{Dan90}, which represent the result in terms of generalized
retarded functions.

\subsection{Local Collision Term and Memory Corrections}\label{LocalCol}

The gradient corrections to the folding of the self-energies with the
propagators in the collision term have already been accounted for in the
kinetic equations. They essentially give rise to in-medium corrections of the
convective part (l.h.s) of the generalized kinetic equation (\ref{keqk}).
Within the spirit of the gradient approximation one also likes to express the
self-energies themselves by space-time local quantities.  Thus, for a
consistent gradient approximation further gradient corrections are admissible
once the self-energy diagrams and thereby the diagrams of $\Phi$ consist of
more than 3 vertices. We call the collision term, evaluated with all Green's
functions in the Wigner representation taken at the same space-time point $X$,
the {\em local} collision term.

In terms of the representation (\ref{Phi-Rmm}) of $\Phi$ and implying
line-sense reversal symmetry for the rates (\ref{Rmm=Rtmm}) the matrix
variation rules (\ref{Coll-var}) determine the following local part of the
collision term (\ref{Coll(kin)}) for a particle of flavor $"a"$ as
%
\def\tp{p'}\def\tm{m'}\def\tW{\widetilde{W}} 
\def\tR{\widetilde{R}}
\begin{eqnarray}
\label{Coll-var-loc} 
C_a^{\scr{loc}} (X,p)=\frac{1}{2}\sum_{m,m'}&&
 \int \dpi{p_1}\cdots\dpi{p_m} \dpi{\tp_1}\cdots\dpi{\tp_{\tm}}
\;R_{m,\tm}(X;p_1,\cdots,p_m;\tp_1,\cdots,\tp_{\tm})
\nonumber \\
&&\times 
\left\{
\Ft_1\cdots\Ft_m \F'_1\cdots\F'_{\tm}
-
\F_1\cdots\F_m \Ft'_1\cdots\Ft'_{\tm} 
\right\}\nonumber\\ 
&&\times
\left[\sum_{i=1}^{m} \delta_{aa_i}\delta^4(p_i-p)
-\sum_{i=1}^{\tm} \delta_{aa_i}\delta^4(\tp_i-p)\right]
\delta^4\left(\sum_{i=1}^{m} p_i - \sum_{i=1}^{\tm} \tp_i \right).
\label{Multi-rate}
\end{eqnarray}
%
The functional variations of $\Phi$ with respect to $\Fd_a$ and $\Fdt_a$ are
expressed in terms of the four-momentum projectors
$\delta_{aa_i}\delta^4(p_i-p)$ which in case of multi-component systems of
particles with different flavors and internal quantum numbers $"a"$ also include
the proper projections onto the different flavors $a_i=a$. This expression
nicely visualizes the detailed balance property namely that the same
multi-particle rate determines both the forward and the backward processes.

The first-order gradient corrections to the local collision term
(\ref{Multi-rate}) we call {\em memory} corrections. We refrain from
specifying the latter in general terms. They are linear in the spatial
gradients of Wigner functions and the displacement factors in front of these
Taylor expansions give momentum gradients of associated other multi-point
functions. An explicit example will be discussed in section
\ref{Memory-Effects}.  Often (but not in general) there are special reasons
why $C_{\scr{mem}}$ is small. In particular, {\em only diagrams of third and
  higher order in the number of vertices give rise to memory effects}, which
are discussed below (sect.  \ref{Memory-Effects}) within a simple model.  One
has to keep in mind that memory effects are quite commonly neglected in
transport models.

\newpage
\subsection{$\Phi$-Derivable Collision Terms} 
\label{Multiprocess-Phi}

\subsubsection{Two-body Potential Interaction}\label{two-body}

To be specific we consider a system of fermions interacting via a two-body
potential $V=V_0 \delta(x-y)$, and, for the sake of simplicity, disregard its
spin structure, by relating spin and anti-symmetrization effects to a
degeneracy factor $d$.  To derive the decomposition of a $\Phi$-derivable
collision term, we employ the rules as described in subsect. \ref{decomp}.

In the first example, we consider the generating functional $\Phi$ to be
approximated by the following two diagrams \vspace*{-6mm}
%
\begin{eqnarray}
\label{Phi-12} 
\ii \Phi =  \frac{1}{2} \loopa + \frac{1}{4} \loopb{c} 
\\ 
\nonumber
\vspace*{16mm} 
\end{eqnarray}
%
with the dashed line illustrating the decomposition according to
(\ref{decomp}). Here the $1/n_\lambda$ factors start with $1/2,1/4,\dots$
according to the nonrelativistic diagram rules for two-body interaction, cf.
Appendix \ref{diagrules}, i.e. the vertex dots are considered as the zero
range limit of a finite range interaction.  In the $\{-+\}$ matrix notation of
the Green's functions, one can easily see that the one-point diagram does not
contribute to the collision term, while decomposing the second diagram along
the dashed line leads to a purely local result
%
\begin{eqnarray}
\label{C2} 
C^{(2)} 
&=& d^2
\int \dpi{p_1} \dpi{p_2} \dpi{p_3}
\left|\;\; \wa{-} \;\; \right|^2
\nonumber 
\\
& \times &
\delta^4\left(p + p_1 - p_2 - p_3\right) 
\left(
\F_2\F_3 \Ft\Ft_1 -
\Ft_2\Ft_3 \F\F_1
\right), 
\end{eqnarray}
%
where the brief notation of the previous subsect. is used for $\F_i$ and
$\Ft_i$. This collision integral has precisely the form of the binary
collision term of Boltzmann--Uehling--Uhlenbeck (BUU), except for the fact
that distribution functions are not constrained to the mass shell. The binary
transition rate
%
\begin{eqnarray}
\label{R2} 
R_2^{(2)} = V_0^2 = \left|\;\; \wa{-} \;\; \right|^2
\end{eqnarray}
%
is nonnegative in this case. Here and below, the bracketed superscript
$^{(2)}$ (or $^{(3)}$) points out the origin of the quantity ($C$, $R$, etc.)
from the second (or third) diagram of $\Phi$. The subscript $_2$ in the
transition rate of eq. (\ref{R2}) indicates the binary-collision nature of
this transition rate.  Note that external 4-momenta (in-going and out-going)
of the scattering amplitude $\wa{-} $ are not confined to the mass shell. For
the trivial case under consideration, this fact does not give rise to any
important consequences. However, for more complicated examples below, it means
that the collision term is determined by off-shell scattering amplitudes.

The picture becomes more complicated, if $\Phi$ involves diagrams of higher
orders. For instance, let us add the following three point diagram to $\Phi$,
i.e.
%
\begin{eqnarray}
\label{Phi-123} 
\ii \Phi &=& \ii\left(\Phi^{(1)} + \Phi^{(2)} + \Phi^{(3)} \right)  
\nonumber 
\\
&=&  \frac{1}{2} \loopa + \frac{1}{4} \loopb{c} 
+ \frac{1}{6} \loopc{c} ,
\\ 
\nonumber
\vspace*{16mm} 
\end{eqnarray}
%
where one possible decomposition is illustrated by dashed lines. The
corresponding self-energy becomes
%
\begin{eqnarray}
\label{Pi-123} 
-\ii \Se &=& -\ii\left(\Se^{(1)} + \Se^{(2)} + \Se^{(3)} \right) = 
\nonumber 
\\
&=&  \pia \; + \;\;\pib \; + \;\;\pic  .
\end{eqnarray}
%
Now the collision term contains a {\em nonlocal} part due to the last
diagram. This nonlocal contribution is discussed in the next subsect. in
detail. The local part can easily be derived in the form
%
\begin{eqnarray}
\label{C30} 
C^{(2)} + C_{\scr{loc}}^{(3)} 
&=& d^2
\int \dpi{p_1} \dpi{p_2} \dpi{p_3}
\left(
\left|\;\; \wa{-} \;\;+ \!\!\!\!\!\wb{-} \;\;\right|^2 - 
\left|\!\!\!\!\!\wb- \;\;\right|^2\right)
\nonumber 
\\[5mm]
&\times &
\delta^4\left(p + p_1 - p_2 - p_3\right) 
\left(
\F_2\F_3 \Ft\Ft_1 -
\Ft_2\Ft_3 \F\F_1
\right),  
\end{eqnarray}
%
where all the vertices in the off-shell scattering amplitudes are of the same
sign, say $"-"$ for definiteness, i.e. there are no $"+-"$ and $"-+"$ Green's
functions left. The quantity $C^{(2)}+C_{\scr{loc}}^{(3)}$ is again of the
Boltzmann form with the transition rate
%
\begin{eqnarray}
\label{R3} 
R^{(2)}_2 +R^{(3)}_2  = 
\left|\;\; \wa{-} \;\; + \!\!\!\!\!\wb{-} \;\;\right|^2 - 
\left|\!\!\!\!\!\wb{-} \;\;\right|^2. 
\end{eqnarray}
%
At the first glance, one may argue that this rate is not necessarily positive
in the limit of strong coupling.  Indeed, the first term in eq.  (\ref{R3}),
i.e. $-\ii V_0$, is purely imaginary, whereas the second one---the
loop---given by
%
\begin{eqnarray}
\label{Loop-A} 
\mbox{loop} = 
-d\int \dpi{p_1}\mid V_0 \mid^2
\ii\Gr^{--}(x,p+p_1 ) \ii\Gr^{--}(x,p_1 ), 
\nonumber
\end{eqnarray}
%
has both real and imaginary parts. Hence, the real part of this loop is
canceled out in eq. (\ref{R3}). If $\mid \Im (\mbox{loop})\mid > \mid V_0
\mid$, the rate $(R_2^{(2)}+R_2^{(3)})$ may become negative, depending on
signs of $V_0$ and $\Im (\mbox{loop})$. However, one has to keep in mind that
the Green's functions in the loop cannot be chosen arbitrarily. Rather in a
consistent treatment, as shown in eq. (80) of ref. \cite{Knoll95}, the loop
reveals a factor 
$\propto\left|1/\Gamma\right|\propto\left|1/V_0^2\right|$ 
resulting from the imaginary part of the retarded self-energy in the
propagators, which balances the total value of the loop. Indeed, in
equilibrium the gain part of collision term and thus $(R_2^{(2)}+R_2^{(3)})$
is positive.  This illustrates that the question of positive definiteness is
quite subtle.

\subsubsection{Bosonization of the Interaction}\label{Bosonization}
 
It is obvious that the situation becomes complicated once one extends the
picture to ring diagrams with more than three vertices.  Yet, there is a
simple (however, not general) strategy to proceed. We may avoid the
two-particle potential interaction from the very beginning and rather
introduce an interaction mediated by an artificially introduced neutral heavy
scalar boson (h.b.) of mass $m_{\scr{h.b.}}$ much larger than any
characteristic momentum transfers in the system. Then the free retarded
Green's function of the boson approximately equals
%
\begin{eqnarray}
\label{delta-b-short}
\Delta^{0 R}_{\scr{h.b.}} \simeq \frac{-1}{m^2_{\scr{h.b.}} -\ii 0}, 
\end{eqnarray} 
%
and the vertex of fermion--heavy-boson interaction being
$g=\sqrt{\left|V_{0}\right|} m_{\scr{h.b.}}$. Moreover, by the same reason the
heavy-boson occupation numbers may be put to zero, $\Delta^{0
  -+}_{\scr{h.b.}}=0$, since typical excitation energies are much less than
the boson mass. The fact that this boson is very heavy makes the
fermion--fermion interaction almost point-like.  To be specific, we assume
that $V_0 <0$, i.e. attractive, which can be mediated by a scalar boson. In
case of repulsive interactions, a vector boson would be an appropriate choice.
  
Thus, {\em{ from now on, we deal with a system of interacting fermions and
    heavy bosons.}}  Let us take the following approximation for the
corresponding $\Phi$ functional (we call it $\Phi_{\scr{h.b.}}$)
\vspace*{-9mm}
%
\begin{eqnarray}
\label{Phi-bos} 
\ii \Phi_{\scr{h.b.}} = \frac{1}{2}\fipi 
\\ 
\nonumber 
\vspace*{26mm} 
\end{eqnarray}
%
in terms of full Green's functions of fermions (the bold solid lines) and
bosons (the bold wavy line). In this approximation, the boson self-energy is
given by (cf. eq. (\ref{varphi}))\footnote{Note that for neutral bosons, which
  number is not conserved, the additional factor $1/2$ appears in eq.
  (\ref{varphi}), cf. ref. \cite{IKV}.}  \vspace*{-5mm}
%
\unitlength1mm
\begin{eqnarray}
\label{Se-bos} 
-\ii \frac{1}{2} \Se_{\scr{h.b.}} = 
\frac{\delta \ii \Phi_{\scr{h.b.}}}{\delta \ii \Gr_{\scr{h.b.}}} = 
\bosprs \hspace*{-14mm}\bosse \hspace*{-3mm}\bosprs
\\ 
\nonumber
\vspace*{26mm} 
\end{eqnarray}
%
and the heavy-boson Green's function is defined by the standard Dyson's
equation \vspace*{-5mm}
%
\begin{eqnarray}
\label{Gr-bos} 
\ii \Gr_{\scr{h.b.}} = \bosprf = \bosprv \hspace*{2mm} + 
\bosprv \hspace*{-3mm}\bosse \hspace*{-3mm}\bosprf .
\\ 
\nonumber
\vspace*{26mm} 
\end{eqnarray}
%
We would like to compare this model with the model described in subsection
\ref{two-body}.  Eliminating the artificial heavy boson one effectively sums
up all ring diagrams of the type\vspace*{-13mm}
%
\begin{eqnarray}
\label{Phi-ring} 
\ii \Phi_{\scr{ring}} =  \frac{1}{2} \loopa + \frac{1}{4} \loopb{ } 
+ \frac{1}{6} \loopc{ } + \frac{1}{8} \fiqa \; + ...
\nonumber
\\ 
\vspace*{29mm} 
\end{eqnarray}
%
Note that $\Phi_{\scr{ring}} \neq \Phi_{\scr{h.b.}}$, since the summations of
loops in eq. (\ref{Phi-bos}) and eq. (\ref{Phi-ring}) have different sense. In
eq. (\ref{Gr-bos}) we have the conventional diagrammatic summation, while in
eq. (\ref{Phi-ring}) the summation is logarithmic, i.e. with the factors
$1/n$, where $n$ is a number of vertices in a diagram. However, this
difference in summations is compensated for by the heavy-boson contribution to
the generating functional (\ref{keediag}). The equivalence of approximations
(\ref{Phi-ring}) and (\ref{Phi-bos}) can be actually seen, e.g., from the fact
that they result in the same approximation for the fermion self-energy
$\ii\Se_{\scr{f}} = \delta \Phi_{\scr{ring}} / \delta \Gr_{\scr{f}} = \delta
\Phi_{\scr{h.b.}} / \delta \Gr_{\scr{f}} $ (cf. eq.  (\ref{varphi})) after
substituting eq. (\ref{Gr-bos}) for the heavy-boson propagator.  Here and
below, the sub-label $"{\scr{f}}"$ denotes fermion quantities.

Upon introduction of the heavy-boson degree of freedom, we have to deal with
two coupled transport equations---for nonrelativistic fermions and for heavy
bosons---with the following collision terms
%
\begin{eqnarray}
\label{C-f} 
C_{\scr{f}} (X,p) 
&=& 
\int \dpi{p_1} \dpi{p_2} g^2
\delta^4\left(p - p_1 - p_2 \right) 
\nonumber 
\\
& \times &
\left[
\Ft_{\scr{f}} (X,p) \F_{\scr{f}} (X,p_1) \F_{\scr{h.b.}} (x,p_2) 
-
\F_{\scr{f}} (X,p) \Ft_{\scr{f}} (X,p_1) \Ft_{\scr{h.b.}} (x,p_2) 
\right], 
\end{eqnarray}
%
for fermions and 
%
\begin{eqnarray}
\label{C-b} 
C_{\scr{h.b.}} (X.p) 
&=& d
\int \dpi{p_1} \dpi{p_2} g^2
\delta^4\left(p + p_1 - p_2 \right) 
\nonumber 
\\
& \times &
\left[
\Ft_{\scr{h.b.}} (X,p) \Ft_{\scr{f}} (X,p_1) \F_{\scr{f}} (X,p_2) 
-
\F_{\scr{h.b.}} (X,p) \F_{\scr{f}} (X,p_1) \Ft_{\scr{f}} (X,p_2) 
\right], 
\end{eqnarray}
%
for heavy bosons, where $g^2 = -V_0 m^2_{\scr{h.b.}}>0$ is defined in terms of
two-particle interaction strength $V_0 <0$ and the heavy boson mass
$m_{\scr{h.b.}}$.

The collision terms (\ref{C-f}) and (\ref{C-b}) are very simple in spite of
the fact that they involve the whole series of ring diagrams. The
corresponding gain and loss terms are positive and contain no memory effects,
as they are hidden in the boson, while in the pure fermionic case already the
triangle diagram gives rise to memory effects.  Indeed, there is no
contradiction here. If one wants to eliminate the bosonic degree of freedom,
one has to resolve the bosonic transport equation with respect to the bosonic
generalized distribution function $\F_{\scr{h.b.}}$ for entire past and
substitute this into the fermionic collision term.  In this way, the resulting
collision term becomes highly complicated and nonlocal and thus contains
memory effects.

Hence, we have demonstrated that sometimes it is useful to introduce new
degrees of freedom in order to achieve a reasonable collision term. Of course,
all the consideration above remains valid also for particle--particle
interaction mediated by a real boson rather than only by artificially
introduced one.

\subsection{Memory Effects in Collision Term}\label{Memory-Effects}

A general treatment of memory effects in the collision term is a cumbersome
task.  In this subsection we continue to consider a system of nonrelativistic
fermions interacting via the contact two-body potential and concentrate on the
third diagram in eq. (\ref{Phi-123}), which has already been considered in a
$\Phi$-derivable scheme for the thermodynamic potential and 
entropy in refs \cite{Riedel,Carneiro}. This is the first ring 
diagram to contribute to memory effects. The corresponding 
self-energy diagram reads 
%
\begin{eqnarray}
\label{Pi-3ij} 
-\ii \Se_{jk}^{(3)}(x,y) &=&  \;\;\pict \;\; , 
\quad j,k \in \{ +,-\} .
\\ 
\nonumber
\vspace*{6mm} 
\end{eqnarray}
%
Standard diagrammatic rules in the matrix representation present
$\Se_{ij}^{(3)}$ in analytic form as
%
\begin{eqnarray}
\label{Pi-3ij-an} 
-\ii\Se_{jk}^{(3)}(x,y) 
= 
\int \di z\cdot  
\ii V_0 \sigma_{jj'}\sigma_{kk'} \ii\Gr^{j'k'}(x,y) 
\Loop^{k'l}(y,z) \sigma_{ll'} \Loop^{l'j'}(z,x), 
\end{eqnarray}
%
where we have introduced a loop function 
%
\begin{eqnarray}
\label{Loop-x} 
\Loop^{jk} (x,y) = \loopl =  d \ii V_0 \ii\Gr^{jk}(x,y) \ii\Gr^{kj}(y,x),
\\ 
\nonumber
\vspace*{10mm} 
\end{eqnarray}
%
and $\sigma_{ij}$ are given by eq. (\ref{sig}).  As above, the factor $d$
results from the trace over spin. In the
Wigner representation, $\Loop^{jk}$ takes the form
%
\begin{eqnarray}\label{Loop-p}
\Loop^{jk}(X, p')= \int \frac{\di^4 p^{\prime\prime}}{(2\pi)^4}
\widetilde{\Loop}^{jk}(X;p^{\prime\prime}+ p',p^{\prime\prime}),
\end{eqnarray}
%
where 
%
\begin{eqnarray}\label{Ltilde}
\widetilde{\Loop}^{jk}(X;p^{\prime\prime}+ p',p^{\prime\prime}) 
= d \ii V_0 \ii\Gr^{jk}(X, p^{\prime\prime}+p') 
\ii\Gr^{kj}(X,p^{\prime\prime}). 
\end{eqnarray}
%
The loop functions $\Loop$ possess notable properties 
%
\begin{eqnarray}
\label{L-prop1}
\Loop^{++} +  \Loop^{--} = \Loop^{-+} +  \Loop^{+-},\,\,\,\,
\Loop^{jk}(x,y) = \Loop^{kj}(y,x),\,\,\,\,\Loop^{jk}(X,p) = \Loop^{kj}(X,-p).
\end{eqnarray}
%
The former property follows from the general property (\ref{Fretarded}) of the
two-point functions and holds in both the coordinate and Wigner
representations.  Proceeding from eq. (\ref{Pi-3ij-an}) and with the help of
relation (\ref{g-rule2}), we can immediately evaluate the Wigner transform of
$\Se_{ij}^{(3)}$ (cf. eq. (\ref{W-transf})) and perform its gradient expansion
%
\begin{eqnarray}
\label{Pi-3ij-exp} 
\Se_{jk}^{(3)} \simeq  \left(\Se_{jk}^{(3)}\right)_{\scr{loc}} + 
\left(\Se_{jk}^{(3)}\right)_{\scr{mem}}, 
\end{eqnarray}
%
where
%
\begin{eqnarray}
\label{Pi-3ij-0} 
-\ii\left(\Se_{jk}^{(3)}\right)_{\scr{loc}}(X,p)= 
\int \frac{\di^4 p'}{(2\pi)^4}   
\ii V_0 \sigma_{jj'}\sigma_{kk'}\sigma_{ll'} \ii\Gr^{j'k'}(X,p'+p) 
\Loop^{k'l}(X,p')  \Loop^{l'j'}(p' , X)
\end{eqnarray}
%
is the local contribution to $\Se_{ij}^{(3)}$, and 
%
\begin{eqnarray}
\label{Pi-3ij-1} 
-\ii\left(\Se_{jk}^{(3)}\right)_{\scr{mem}}(X,p) &=& 
\frac{\ii}{2}
\int \frac{\di^4 p'}{(2\pi)^4}   
\ii V_0 \sigma_{jj'}\sigma_{kk'}\sigma_{ll'} \ii\Gr^{j'k'}(X,p'+p) 
\nonumber \\ 
&\times& 
\Pbr{\Loop^{k'l}(X,p'), \Loop^{l'j'}(X,p')}_{p' ,X} 
\end{eqnarray}
%
is the first-order gradient (memory) correction, which is of central interest
in this subsection. Here, the Poisson bracket is taken with respect to the
$(p',X)$ variables indicated in the subscript.

To evaluate the memory correction for the collision term (cf. eqs (\ref{gain})
and (\ref{Coll(kin)})), we need to consider the following combinations of
self-energies with Green's functions
%
\begin{eqnarray}
\label{loss-3} 
\left(\Se_{+-}^{(3)}\right)_{\scr{mem}}(X,p) \Gr^{-+}(X,p) = \frac{\ii}{2}
\int \frac{\di^4 p'}{(2\pi)^4}   
\frac{1}{d}\widetilde{\Loop}^{+-}(X;p'+p,p) 
\sigma_{ll'} \Pbr{\Loop^{-l}, \Loop^{l'+}}_{p' ,X}\, , 
\end{eqnarray}
%
%
\begin{eqnarray}
\label{gain-3} 
\Gr^{+-}(x,p) \left(\Se_{-+}^{(3)}\right)_{\scr{mem}}(X,p) = \frac{\ii}{2}
\int \frac{\di^4 p'}{(2\pi)^4}   
\frac{1}{d}\widetilde{\Loop}^{-+}(X;p'+p,p) 
\sigma_{ll'} \Pbr{\Loop^{+l}, \Loop^{l'-}}_{p' , X}\, . 
\end{eqnarray}
%
With the symmetry relations  
%
\begin{equation}
\label{pois-rel}
\sigma_{ll'} \Pbr{\Loop^{-l}, \Loop^{l'+}}_{p' , X} = 
\sigma_{ll'} \Pbr{\Loop^{+l}, \Loop^{l'-}}_{p' ,X} =
\Pbr{\Loop^{+-}, \Loop^{-+}}_{p' , X} 
\end{equation}
%
deduced from (\ref{L-prop1}) one determines the first-order gradient
correction to the collision term induced by graph (\ref{Pi-3ij}) as
%
\begin{eqnarray}
\label{col-pi-delta}
&&C_{\scr{mem}}^{(3)}(X,p) = 
\left[
\left(\Se_{+-}^{(3)}\right)_{\scr{mem}}(X,p)\Gr^{-+}(X,p) 
-
\Gr^{+-}(X,p) \left(\Se_{-+}^{(3)}\right)_{\scr{mem}}(X,p)
\right] 
\nonumber \\
&&=
\frac{\ii}{2}
\int \frac{\di^4 p'}{(2\pi)^4}   
\frac{1}{d}\left[
\widetilde{\Loop}^{+-}(X;p'+p,p) - \widetilde{\Loop}^{-+}(X;p'+p,p) 
\right] 
\Pbr{\Loop^{+-}, \Loop^{-+}}_{p' ,X}\, .
\end{eqnarray}
%
The corresponding local collision term is given by eq. (\ref{C30}).

\section{H--Theorem} \label{H-theorem}

\subsection{Time-Irreversibility of the Generalized Kinetic Description}

Compared to exact descriptions, which are time reversible, reduced description
schemes in terms of relevant degrees of freedom have access only to some
limited information and thus normally lead to irreversibility. Various
reduction and coarse graining schemes have been discussed and developed over
the years. In the Green's function formalism presented here the information
loss arises from the truncation of the exact Martin--Schwinger
hierarchy, 
where the exact one-particle Green's function couples to the
two-particle Green's functions, cf.  \cite{Kad62,Bot90}, which in turn are coupled
to the three-particle level, etc. \footnote{The appropriate classical limit
  leads to the Bogolyubov--Born--Green--Kirkwood--Yvon chain of equations.}
This truncation is achieved by the standard Wick decomposition, where all
observables are expressed through one-particle propagators and therefore
higher-order correlations are dropped. This step provides the Dyson equation
(\ref{Dyson}) and the corresponding loss of information is expected to lead to
a growth of entropy with time.

At the operator level the nonequilibrium entropy can be straightforwardly 
formulated in terms of von Neumann's entropy, which is an entropy in the
information theory sense \cite{AKatz}
%
\begin{eqnarray}\label{vNeumann}
S=-\Tr \medhat{\rho}\ln\medhat{\rho}.
\end{eqnarray}
%
It is given by the expectation value of the logarithm of the density operator
$\medhat{\rho}$ itself. The problem of the Green's function formalism is that
it does not give a direct access to the density operator itself but rather
describes the space-time dependence of the expectation values of well defined
operators, say $\left<\medhat{A(t)}\right>=\Tr\medhat{A}\medhat{\rho}(t)$. To
this extend, there is no immediate formulation of the entropy (\ref{vNeumann})
in terms of Green's functions for nonequilibrium case. The situation is
different at equilibrium, where with
$\medhat{\rho}=\exp(-\beta(\medhat{H}-\mu\medhat{N}))/Z$ a well defined
density operator exists, which leads to the well known Matsubara or real-time
formulation of the equilibrium entropy, cf. sect. \ref{Ther-entr} below. As
will be seen, even in equilibrium, the entropy expression is not priori given
but rather depends on the choice of $\Phi$ in a $\Phi$-derivable scheme.

In order to access a nonequilibrium expression relevant for our generalized
transport equation (\ref{keqk}), we shall start from this transport equation
and derive a flow expression $s^\mu$ with the property that its divergence
grows in time, i.e.
%
\begin{eqnarray}\label{dmusmu>0}
\partial_\mu s^\mu(X)\ge 0,
\end{eqnarray}
%
and which in the equilibrium limit merges the corresponding equilibrium form
of the entropy flow. That is, we show the existence of an $H$-theorem for our
generalized kinetic description and $s^\mu(X)$ is thus identified with the
kinetic entropy flow. Thereby, the validity conditions for the derivation of
this kinetic entropy flow coincide with those of the kinetic equations
themselves.

\subsection{Markovian Entropy Flow}\label{Flow}

We start with general manipulations which lead us to definition of the kinetic
entropy flow.  We multiply eq. (\ref{keqkt}) by $(\mp)\ln(\Ft/A)$, eq.
(\ref{keqk}) by $-\ln(\F/A)$, take their sum, integrate it over $\di^4 p/(2\pi
)^4$ and finally sum the result over internal degrees of freedom like spin
($\Tr$). Using the identity for the Poisson brackets
%
\begin{eqnarray}\label{BbrlnF}
\Pbr{B,A f}\ln f\pm\Pbr{B,A(1\mp f)}\ln (1\mp f)
=\Pbr{B,A f\ln f\pm A(1\mp f)\ln (1\mp f)}
\end{eqnarray}
%
for any functions $A,B$ and positive $f$ and $1\mp f$, one then arrives at the
following relation
%
\begin{eqnarray}
\label{s-eq.} 
\partial_\mu s^\mu_{\scr{loc}} (X) =\mbox{Tr}\sum_a
\int \dpi{p} \ln \frac{\Ft_a}{\F_a} C_a (X,p),   
\end{eqnarray}
%
where the quantity 
%
\begin{eqnarray}
\label{S(gen)} 
s^\mu_{\scr{loc}}  = \sum_a s^\mu_{\scr{loc},a} =\mbox{Tr}
\sum_a\int \dpi{p}
\left[
\left(
\vu^{\mu} - \frac{\partial \Re\Se^R_a}{\partial p_{\mu}}
\right)
\left(
\mp \Ft_a \ln \frac{\Ft_a}{A_a} - \F_a \ln \frac{\F_a}{A_a} 
\right)
\right.\;\,&&
\nonumber 
\\
- 
\left. 
\Re\Gr^R_a
\left(
\mp \ln\frac{\Ft_a}{A_a} \frac{\partial}{\partial p_{\mu}} 
\left(\Gm_a\frac{\Ft_a}{\A_a} \right)
- 
\ln\frac{\F_a}{A_a} \frac{\partial}{\partial p_{\mu}} 
\left(\Gm_a\frac{\F_a}{\A_a} \right)
\right)
\right]&& 
\end{eqnarray}
%
obtained from the l.h.s. of the kinetic equation is interpreted as the local
(Markovian) part of the entropy flow. Here we have restored the summation over
$"a"$ denoting the different particle species and intrinsic quantum numbers for
a multi-component system. It illustrates that, although this entropy
expression accounts for interactions among all particles, it can be expressed
as a sum of the individual contributions, each of which is solely determined
by the particle self-energy $\Re\Se^R_a$ and its width $\Gm_a$. Gradient
corrections to the collision term (i.e.  $C_{\scr{mem}}$) give rise to extra
memory contributions to the entropy flow.

Partial integrations in eq. (\ref{S(gen)}) lead us to a more transparent
expression for the entropy flow
\begin{eqnarray}\label{entr-transp}
s^\mu_{\scr{loc}}  = \mbox{Tr}\sum_a\int \dpi{p} 
A^{\mu}_{sa} (X,p)\sigma_a (X,p),\,\,\, 
\sigma_a (X,p) 
=\mp [1\mp f]\ln [1\mp f]-f\ln f ,
\end{eqnarray}
where 
\begin{eqnarray}\label{As}
A^{\mu}_{sa} (X,p)=\frac{A_a \Gamma_a }{2}B^{\mu}_{a}
\end{eqnarray}
has the meaning of an entropy-flow spectral function, while
\begin{eqnarray}\label{B-j}
B^{\mu}_{a} =A_a \left[
\left(
\vu^{\mu} - \frac{\partial \Re\Sa^R }{\partial p_\mu}
\right)
- M_a \Gamma_a^{-1}\frac{\partial \Gamma_a}{\partial p_\mu}
\right], 
\end{eqnarray}
is the flow spectral function, cf. the corresponding drift term (proportional
to $\partial_{\mu}f$ in eq. (\ref{keqk2})). The zero-components of these
functions coincide with the corresponding two functions introduced in refs
\cite{Weinhold,Weinhold-th} for the case of equilibrium systems. Moreover, the
$B^0$ function satisfies the sum rule
\begin{eqnarray}
\int_{-\infty}^{\infty}\frac{\di p_0}{2\pi}B^0_{a}=1,
\end{eqnarray}
which is easily obtained from the sum rule (\ref{A-sumf}) for the spectral
function $A$ with the help of Kramers--Kronig dispersion relations. For the
case that the considered particle $a$ is a resonance, like the $\Delta$ or
$\rho$-meson resonances in hadron physics, the $B^0$ function relates to the
energy variations of scattering phase shift of the scattering channel coupling
to the resonance in the virial limit, for details see e.g.  refs
\cite{Weinhold,Weinhold-th,IKV}.

In the noninteracting-particle limit, the entropy (the zero component of
$s^\mu_{\scr{loc}}$) directly transforms into the proper ideal gas expression,
cf. ref. \cite{Land80}.  In the quasiparticle approximation, the $\Ldt$ and
$\Ld$ terms have an additional smallness, which allows to neglect these terms.
Thus, expression (\ref{S(gen)}) for the entropy flow takes the form
%
\begin{eqnarray}
\label{S(qp)} 
\!\!\!\!\!\!\!\!
(s_{\scr{loc}}^\mu)^{\scr{qp}}  = \left(
\begin{array}{cc}
&s_0\\
&\vec s 
\end{array} \right)
=\mbox{Tr} 
\sum_a \int \frac{\di^3 p}{(2\pi)^3}
\left(
\begin{array}{cc}
&1\\
&\displaystyle
\frac{\partial \varepsilon_a}{\partial {\vec p}}
\end{array} 
\right)
\left[ \vphantom{\int}
\mp\left(1\mp f^{\scr{qp}}_a\right) 
\ln \left(1\mp f^{\scr{qp}}_a\right) 
-
f^{\scr{qp}}_a \ln f^{\scr{qp}}_a 
\right] 
\end{eqnarray}
%
in the quasiparticle limit, which follows from the substitution of eqs
(\ref{F-q.p.}) and (\ref{Ft-q.p.})  into eq. (\ref{S(gen)}). From
(\ref{entr-transp}) eq. (\ref{S(qp)}) is also easily recovered, since $A^2
\Gamma /2$ transforms to the corresponding $\delta$-function in the limit
$\Gamma \rightarrow 0$.

To prove explicitly the $H$-theorem we have to show that the r.h.s. of eq.
(\ref{s-eq.}) is nonnegative. To this end, we should consider the convolution
of the collision term with $\ln (\Ft/\F)$.  First, we do this for collision
terms in local approximation.

\subsection{Local Collision Term and H-Theorem} 
\label{ECT}

Using the multi-particle process decomposition (\ref{Multi-rate}) we arrive at
the relation
%
\begin{eqnarray}
\label{s(coll)} 
&&\mbox{Tr}\int \dpi{p} \ln \frac{\Ft}{\F} C_{\scr{loc}} (X,p) 
= \mbox{Tr}
\sum_{m,\tm}
\frac{1}{2} 
 \int \dpi{p_1}\cdots\dpi{p_m} \dpi{\tp_1}\cdots\dpi{\tp_{\tm}}
\nonumber 
\\
&&\times
\left\{\F_1\cdots\F_m \Ft'_1\cdots\Ft'_{\tm} 
-
\Ft_1\cdots\Ft_m \F'_1\cdots\F'_{\tm}\right\}
\ln\frac{\F_1 \cdots\F_m \Ft'_1\cdots\Ft'_{\tm}}
        {\Ft_1\cdots\Ft_m\F'_1 \cdots\F'_{\tm}}
\nonumber 
\\
&&\times
R_{m,\tm}\; 
\delta^4\left(\sum_{i=1}^{m} p_i - \sum_{i=1}^{\tm} \tp_i \right).
\end{eqnarray}
%
Here we assumed different flavors and intrinsic quantum number to be absorbed
in the momenta $p_1$ and $\tp_i$. In the case when all rates $R_{m,\tm}$ are
nonnegative, i.e. $R_{m,\tm}\ge 0$, this expression is nonnegative, since
$(x-y)\ln(x/y) \ge 0$ for any positive $x$ and $y$.  In particular,
$R_{m,\tm}\ge 0$ takes place for all $\Phi$-functionals up to two vertices.
Then the divergence of $s_{\scr{loc}}^\mu$ is nonnegative which proves the
$H$-theorem in this case with (\ref{S(gen)}) as the nonequilibrium entropy
flow.

\subsection{Explicit examples for the H-Theorem} 
\label{DCT}

We explicitly discuss the two examples introduced already in sect.
\ref{Multiprocess-Phi} with Markovian collision terms, i.e. with the $\Phi$
functional consisting of one- and two-point diagrams only. In the pure
fermionic case with collision term (\ref{Phi-12}), one can state an exact
$H$-theorem
%
\begin{eqnarray}
\label{s2(coll)} 
&&\mbox{Tr}\int \dpi{p} \ln \frac{\Ft}{\F} C^{(2)} 
=
d^3
\int \dpi{p} \dpi{p_1} \dpi{p_2} \dpi{p_3}
\left|\;\; \wa \;\; \right|^2
\nonumber 
\\
& \times &
\delta^4\left(p + p_1 - p_2 - p_3\right) 
\ln \frac{\Ft\Ft_1\F_2\F_3}{\F\F_1\Ft_2\Ft_3}
\left(
\Ft\Ft_1 \F_2\F_3 -
\F\F_1   \Ft_2\Ft_3 
\right) \ge 0. 
\end{eqnarray}
%
Note that here and below the extra factor $d$ comes from summation ($\Tr$)
over spin of fermions.

Furthermore, it is instructive to consider the $\Phi$ approximation
(\ref{Phi-bos}), where a heavy scalar boson has been introduced in order to
sum up the series of ring diagrams of eq. (\ref{Phi-ring}), cf.  subsect.
\ref{Bosonization}.  In this case, there are two coupled kinetic equations
with collision terms (\ref{C-f}) for fermions and (\ref{C-b}) for heavy
bosons, leading to
%
\begin{eqnarray}
\label{s(fb)-eq.} 
\partial_\mu s^{\mu}_{\scr{loc}}=\partial_\mu s^{\mu}_{\scr{f}}+
\partial_\mu s^{\mu}_{\scr{h.b.}} =
\int \dpi{p} 
\left( d     \ln \frac{\Ft_{\scr{f}}}{\F_{\scr{f}}} C_{\scr{f}} + 
\frac{1}{2} \ln \frac{\Ft_{\scr{h.b.}}}{\F_{\scr{h.b.}}} C_{\scr{h.b.}}
\right),   
\end{eqnarray}
%
where $s^\mu_{\scr{loc}}$ is given by the sum of the proper fermion
($s^{\mu}_{\scr{f}}$) and heavy-boson ($s^{\mu}_{\scr{h.b.}}$) contributions.
The thermodynamic entropy for this system, i.e. for $\Phi$ given by diagram
(\ref{Phi-bos}), has recently been obtained in ref. \cite{Vanderheyden}.  Here
we present the corresponding nonequilibrium entropy flow together with an
affirmation of the $H$-theorem also for this case. The collision term, indeed,
becomes
%
\begin{eqnarray}
\label{H(fb)-theorem} 
\partial_\mu s^\mu_{\scr{loc}} 
&=& d
\int \dpi{p_1} \dpi{p_2} \dpi{p_b} g^2
\delta^4\left(p_1 - p_2 - p_b \right)\nonumber \\ 
&\times &\ln \frac{\Ft_{1} \F_{2} \F_{\scr{h.b.}}}%
{\F_{{\scr{f}}1} \Ft_{{\scr{f}}2} \Ft_{\scr{h.b.}}}
\left(\Ft_{{\scr{f}}1} \F_{{\scr{f}}2} \F_{\scr{h.b.}} 
-
\F_{{\scr{f}}1} \Ft_{{\scr{f}}2} \Ft_{\scr{h.b.}} 
\right) \ge 0, 
\end{eqnarray}
%
which is nonnegative.

Our representation of the entropy of a system interacting via two-body
potential (i.e. as a sum of a purely fermionic part and that of the
artificially introduced heavy boson) is also very similar to that derived by
Riedel \cite{Riedel} within the ring-diagram model of $\Phi$-derivable
thermodynamics. In both cases, the bosonic part of the entropy
$s^\mu_{\scr{h.b.}}$ takes account of the fermion--fermion interaction
calculated within the ring-diagram approximation (\ref{Phi-ring}). In
thermodynamics, this interaction part of the entropy gives rise to the famous
correction to the specific heat of liquid $^3$He
\cite{Riedel,Carneiro,Baym91}: $\sim T^3 \ln T$, where $T$ is the temperature.
As has been found by Carneiro and Pethick \cite{Carneiro}, this correction to
the specific heat emerges already solely from the third diagram of the whole
ring series (\ref{Phi-ring}). To demonstrate the same within our kinetic
approach, we should consider the $\Phi$-derivable model (\ref{Phi-123})
involving only the first three ring diagrams.  Moreover, since the local
entropy expression (cf. eq. (\ref{S(gen)})) derived above does not contain
such kind of corrections, one has to explicitly consider memory effects in the
collision term (\ref{col-pi-delta}).

\subsection{Memory Effects in Entropy Flow and H-Theorem} 
\label{Memory in Entropy}

We assume that the fermion--fermion potential interaction is such that the
corresponding transition rate (\ref{R3}) is always nonnegative, so that the
$H$-theorem takes place in the local approximation, i.e. when we keep only
$C^{(2)}+C_{\scr{loc}}^{(3)}$. Our aim now is to derive the entropy, which
takes into account memory effects in the collision term
($C_{\scr{mem}}^{(3)}$).

Proceeding similarly to that in subsect. \ref{Flow}, we multiply eq.
(\ref{keqkt}) by $-\ln(\Ft/A)$, eq. (\ref{keqk}) by $-\ln(\F/A)$, make their
sum and integrate it over $\di^4 p/(2\pi )^4$. Thus, we arrive at the equation
%
\begin{equation}
\label{log-term}
\partial_\mu s^\mu_{\scr{loc}} (X) =\mbox{Tr}
\int \dpi{p} \ln \frac{\Ft}{\F} (C^{(2)}+C_{\scr{loc}}^{(3)})
+\mbox{Tr}\int \dpi{p} \ln \frac{\Ft}{\F} C_{\scr{mem}}^{(3)},    
\end{equation}
%
where $ s^\mu_{\scr{loc}}$ is still the Markovian entropy flow defined by eq.
(\ref{S(gen)}). Our aim here is to present the last term on the r.h.s. of eq.
(\ref{log-term}) in the form of full $x$-derivative
%
\begin{equation}
\label{ent-mar}
\mbox{Tr}\int \dpi{p} \ln \frac{\Ft}{\F} C_{\scr{mem}}^{(3)} 
=-\partial_{\mu} s_{\scr{mem}}^{\mu} (X) + \delta c_{\scr{mem}} (X)
\end{equation}
%
of some function $s_{\scr{mem}}^{\mu} (X)$, which we then interpret as a
non-Markovian correction to the entropy flow of eq. (\ref{S(gen)}), plus a
correction ($\delta c_{\scr{mem}}$) which is small in some sense.  Indeed,
this term on the r.h.s. of eq. (\ref{ent-mar}) is linear in $X$- and
$p$-derivatives. Hence, it cannot be transformed into sign-definite form. The
only possibility remained is to construct a full derivative of it. If we
succeed to find a proper $s_{\scr{mem}}^{\mu} (X)$, then relying on smallness
of $\delta c_{\scr{mem}}$ we obtain
%
\begin{equation}
\label{H-der}
\partial_\mu \left(s^\mu_{\scr{loc}} + s_{\scr{mem}}^{\mu}\right) \simeq
\mbox{Tr}\int \dpi{p} \ln \frac{\Ft}{\F} (C^{(2)}+C_{\scr{loc}}^{(3)}) \geq 0, 
\end{equation}
%
which is the $H$-theorem for the non-Markovian kinetic equation under
consideration with $s^\mu_{\scr{loc}} + s_{\scr{mem}}^{\mu}$ as the proper
entropy flow. The r.h.s. of eq. (\ref{H-der}) is nonnegative due to our
assumption that the corresponding transition rate (\ref{R3}) is always
nonnegative.

Hence, let us consider the last term on the r.h.s. of eq. (\ref{log-term}).
Upon substituting expression (\ref{col-pi-delta}) for $C_{\scr{mem}}^{(3)}$
and shifting the integration variable $p\rightarrow p-p'/2$ in this term, we
arrive at
%
\begin{eqnarray}\label{I}
\hspace*{-10mm}
\mbox{Tr}\int \dpi{p} \ln \frac{\Ft}{\F} C_{\scr{mem}}^{(3)}&=&
\frac{\ii}{2}
\int \frac{ \di^4 p }{(2\pi)^4}
\frac{ \di^4 p'}{(2\pi)^4}
\left[\widetilde{\Loop}^{+-}(X;p+ \frac{p'}{2} , p- \frac{p'}{2})- 
\widetilde{\Loop}^{-+}(X;p+ \frac{p'}{2} , p- \frac{p'}{2})
\right]\nonumber \\
&\times&
\Pbr{ \Loop^{+-}(X,p'), \Loop^{-+}(X, p')}_{p',X}
\ln\frac{\widetilde{F}(X,p- p'/2)}{F(X,p-p'/2)}.
\end{eqnarray}
%
By making the half-sum of eq. (\ref{I}) with that with replaced $p'\rightarrow
-p'$, we arrive at the symmetric form of this equation
%
\begin{eqnarray}\label{Isym}
\hspace*{-10mm}
\mbox{Tr}\int \dpi{p} \ln \frac{\Ft}{\F} C_{\scr{mem}}^{(3)}&=&
\frac{\ii}{4}
\int \frac{ \di^4 p }{(2\pi)^4}
\frac{ \di^4 p'}{(2\pi)^4}
\left[
\widetilde{\Loop}^{-+}(X;p+ \frac{p'}{2} , p- \frac{p'}{2})-
\widetilde{\Loop}^{+-}(X;p+ \frac{p'}{2} , p- \frac{p'}{2}) 
\right]
\nonumber \\
&\times&
\{ \Loop^{-+}(X, p'), \Loop^{+-}(X,p')\}_{p',X}
\mbox{ln}\frac{\widetilde{\Loop}^{+-}(X;p+p'/2,p-p'/2)}
{\widetilde{\Loop}^{-+}(X;p+p'/2,p-p'/2)}, 
\end{eqnarray}
%
which after simple algebraic transformations can be decomposed into two terms
%
\begin{equation}\label{I-deltaI}
\mbox{Tr}\int \dpi{p} \ln \frac{\Ft}{\F} C_{\scr{mem}}^{(3)}
=c_{\scr{mem}}+\delta c_{\scr{mem}} ,
\end{equation}
%
with
%
\begin{eqnarray}\label{Isym0}
c_{\scr{mem}}(X)&=& 
\frac{\ii}{2}
\int \frac{ \di^4 p }{(2\pi)^4}
\frac{ \di^4 p'}{(2\pi)^4}
\left\{ \vphantom{\frac{\widetilde{\Loop}^{+-}}{\widetilde{\Loop}^{-+}}}
\Loop^{-+}(X,p'), \Loop^{+-}(X, p')
\widetilde{\Loop}^{-+}\left(X;p+\frac{p'}{2},p-\frac{p'}{2}\right)
\right.
\nonumber \\ 
&\times& \left.
\left[
\mbox{ln}\frac{\widetilde{\Loop}^{+-}(X;p+p'/2,p-p'/2)}%
{\widetilde{\Loop}^{-+}(X;p+p'/2,p-p'/2)}-1\right]
\right\}_{p',X},
\end{eqnarray}
%
%
\begin{eqnarray}\label{Isym-delta}
\delta c_{\scr{mem}}(X)&=& 
\frac{\ii}{2}
\int \frac{ \di^4 p }{(2\pi)^4}
\frac{ \di^4 p'}{(2\pi)^4}
\left[
\left\{ 
\Loop^{-+}(X,p'), \Loop^{+-}(X, p')\right\}_{p',X}
\widetilde{\Loop}^{-+}\left(X;p+\frac{p'}{2},p-\frac{p'}{2}\right)
\right.
\nonumber \\ 
&-& \left.
\left\{ 
\vphantom{\frac{\widetilde{\Loop}^{+-}}{\widetilde{\Loop}^{-+}}}
\Loop^{-+}(X,p'), 
\widetilde{\Loop}^{-+}\left(X;p+\frac{p'}{2},p-\frac{p'}{2}\right)
\ln\frac{\widetilde{\Loop}^{+-}(X;p+p'/2,p-p'/2)}%
{\widetilde{\Loop}^{-+}(X;p+p'/2,p-p'/2)}
\right\}_{p',X}
\right.
\nonumber \\ 
&\times& \left.
\vphantom{\frac{\widetilde{\Loop}^{+-}}{\widetilde{\Loop}^{-+}}}
\Loop^{+-}(X, p')\right].
\end{eqnarray}
%
Using partial integration, we have subdivided the quantity of eq.
(\ref{I-deltaI}) in such a way that the first term $c_{\scr{mem}}$ takes the
form of the full divergence, and thus defines the non-Markovian contribution
to the entropy flow (cf. eq. (\ref{ent-mar}))
%
\begin{eqnarray}\label{entr-flow-Mark}
s_{\scr{mem}}^{\nu}(X) &=& - \frac{\ii}{2}
\int \frac{ \di^4 p }{(2\pi)^4}\
\frac{ \di^4 p'}{(2\pi)^4}
\widetilde{\Loop}^{-+}\left(X;p+\frac{p'}{2},p-\frac{p'}{2}\right)
\Loop^{+-}(X,p')
\nonumber \\
&\times&
\left[
\ln\frac{\widetilde{\Loop}^{+-}(X;p+p'/2,p-p'/2)}%
{\widetilde{\Loop}^{-+}(X;p+p'/2,p-p'/2)}-1\right]
\frac{\partial \Loop^{-+}(X,p')}{\partial p'_\nu}.  
\end{eqnarray}
%
This $c_{\scr{mem}}$-term remains nonzero even in the local thermal
equilibrium, which is defined by the former equilibrium relations 
(\ref{KMS-G})--(\ref{occup}) but with temperature $T(X)$, 4-velocity 
$U^\nu(X)$ and chemical potential $\mu(X)$ depending on the
coordinates. On the contrary, as we show below, the second term
$\delta c_{\scr{mem}}$ vanishes in the limit of local thermal equilibrium. 

In local thermal equilibrium the Kubo-Martin-Schwinger condition (\ref{KMS-G})
provides the following relations
%
\begin{equation}\label{loc--log}
\left(\frac{\widetilde{\Loop}^{+-}(X;p+p'/2,p-p'/2)}%
{\widetilde{\Loop}^{-+}(X;p+p'/2,p-p'/2)}\right)_{\scr{loc.eq.}}
=\left(\frac{\Loop^{+-}(X;p')}{\Loop^{-+}(X,p')}\right)_{\scr{loc.eq.}}
=\exp\left(
\frac{p'_\nu U^\nu(X)}{T(X)} 
\right),
\end{equation}
%
which can also be derived from (\ref{Geq})--(\ref{fb-occup}) proceeding from
definitions of $\Loop^{ij}$ (\ref{Loop-p}) and $\widetilde{\Loop}^{ij}$
(\ref{Ltilde}). Guided by (\ref{loc--log}) we write the
$\widetilde{\Loop}^{ij}$ ratio in the $\ln$-term of (\ref{Isym-delta}) as
%
\begin{equation}\label{loc--log+x}
\frac{\widetilde{\Loop}^{+-}(X;p+p'/2,p-p'/2)}%
{\widetilde{\Loop}^{-+}(X;p+p'/2,p-p'/2)}
=\frac{\Loop^{+-}(X;p')}{\Loop^{-+}(X,p')} (1+\xi), 
\end{equation}
%
where $\xi$ is expected to be small within the validity range of the
generalized kinetic equation (\ref{keqk}), i.e. $|\xi|\approx|f-\gamma|$.
Substituting this into expression (\ref{Isym-delta}) for $\delta
c_{\scr{mem}}$, the $\ln(\Loop^{-+}/\Loop^{-+})$-term in the second Poisson
bracket cancels against the first term, since the $p$-integration converts the
linear $\widetilde{\Loop}^{ij}$ factor into $\Loop^{ij}$ factor. Thus one
obtains
%
\begin{eqnarray}\label{delta-D-x}
\delta c_{\scr{mem}}&=& 
-\frac{\ii}{2}
\int \frac{ \di^4 p }{(2\pi)^4}
\frac{ \di^4 p'}{(2\pi)^4}
\left\{ 
\Loop^{-+}, 
\widetilde{\Loop}^{-+} 
\ln(1+\xi)\right\}_{p',X}
\Loop^{+-} 
\nonumber \\ 
&\simeq& 
\frac{\ii}{2}
\int \frac{\di^4 p }{(2\pi)^4}
\frac{\di^4 p'}{(2\pi)^4}
\left\{ 
\Loop^{-+}, 
\widetilde{\Loop}^{-+} 
\frac{1}{2} \xi^2\right\}_{p',X}
\Loop^{+-} 
,
\end{eqnarray}
%
where also the term linear in $\xi$ exactly cancels out, since
%
\begin{eqnarray}\label{delta-D-x1}
&&-\frac{\ii}{2}
\int \frac{ \di^4 p }{(2\pi)^4}
\frac{ \di^4 p'}{(2\pi)^4}
\left\{ 
\Loop^{-+}, 
\widetilde{\Loop}^{-+} 
\xi\right\}_{p',X}
\Loop^{+-} 
\nonumber \\ 
&=& 
-\frac{\ii}{2}
\int \frac{ \di^4 p }{(2\pi)^4}
\frac{ \di^4 p'}{(2\pi)^4}
\left\{ 
\Loop^{-+}, 
\left(\frac{\Loop^{-+}}{\Loop^{+-}}\widetilde{\Loop}^{+-} 
- \widetilde{\Loop}^{-+}\right) 
\right\}_{p',X}
\Loop^{+-} 
=0.
\end{eqnarray}
%
As above, the $\widetilde{\Loop}^{ij}$ can be readily integrated over $p$ to
produce $\Loop^{ij}$, revealing a cancellation of the terms in the Poisson
bracket in (\ref{delta-D-x1}).  Thus, $\delta c_{\scr{mem}}$ is not only zero
in local equilibrium ($\xi=0$). It is of the second-order in the small
parameter $|\xi|\approx|f-\gamma|$ times gradients and therefore negligible
compared to $c_{mem}$ in (\ref{Isym0}) which is of linear order in
$|\xi|\approx|f-\gamma|$ times gradients. Thus within the validity of the
generalized kinetic equation (\ref{keqk}) $s_{\scr{mem}}^{\nu}(X)$ as given in
(\ref{entr-flow-Mark}) represents the appropriate non-Markovian memory
correction to the entropy flow.  In local equilibrium expression this term can
be further simplified to
%
\begin{equation}\label{s3-Mark}
(s_{\scr{mem}}^{\nu})_{\scr{eq}}=
\frac{\ii}{2}
\int 
\frac{ \di^4 p'}{(2\pi)^4}
\Loop^{-+}(X;p')\Loop^{+-}(X,p')
\left[
\mbox{ln}\frac{\Loop^{+-}(X,p')}{\Loop^{-+}(X,
p')}-1\right]\frac{\partial \Loop^{-+}(X,p')}{\partial
p'_{\nu}} 
\end{equation}
%
by means of relation (\ref{loc--log}).

\section{Thermodynamic Limit of Entropy}\label{Ther-entr}

\subsection{Thermodynamic Entropy}\label{Ther-rel}

In the Matsubara technique, the thermodynamic potential $\Omega$ (see, e.g.,
ref. \cite{Abrikos}) is a functional of Matsubara Green's functions $\Gr
(\varepsilon_n , \vec p )$. Its important property is that it is stationary
under variations of $\Gr (\varepsilon_n , \vec p )$ at fixed free Matsubara
Green's function $\Gr^0(\varepsilon_n ,\vec p )$
%
\begin{equation}\label{st-prop}
\left(\frac{\delta \Omega}{\delta \Gr(\varepsilon_n ,\vec p )}
\right)_{\Gr^0}=0.
\end{equation}
%
Since $\Gr(\varepsilon_n ,\vec p )$ has the spectral representation
%
\begin{equation}\label{spec-maz}
\Gr(\varepsilon_n ,\vec p )=\int_{-\infty}^{\infty}
\frac{\di\varepsilon'}{2\pi}\frac{A(\varepsilon',\vec p)}%
{\varepsilon_n -\varepsilon'} ,
\end{equation}
%
property (\ref{st-prop}) implies that $\Omega$ is stationary under variations
of the spectral density $A(\varepsilon,\vec p)$, provided the Matsubara
frequencies $\varepsilon_n $, over which the summation runs, are not changed
and, thus, the free Matsubara Green's function $\Gr^0(\varepsilon_n ,\vec p )$
is unaltered.

In the standard way, cf. (\ref{Real-Mats})--(\ref{Mats-spectral}), the
$\varepsilon_n$-sum is converted into an energy integral over the distribution
functions $n(\varepsilon)$, cf. (\ref{occup}), and thermodynamic potential
expressed in terms of the real-time Green's functions and self-energies
becomes \cite{Carneiro} (cf. eqs. (\ref{Geq}) and (\ref{Seq})),
%
\begin{eqnarray}
\label{Om-maz-cont} 
\Omega =
\mbox{Tr} \int \di^3 x \dpi{ p} n (\varepsilon)
\left(-2\mbox{Im} \;
\ln\left[-\Gr^R(p_0 +\ii 0 ,\vec p )\right]
-
 \Re\Gr^R\Gamma -A\Re\Se^R\right)
+\Phi_T 
\end{eqnarray}  
%
where $\varepsilon = p_0 -\mu$ in the rest frame of the equilibrated system,
and $\Phi_T$ is represented by the same set of closed diagrams as $\Phi$.  The
thermodynamic entropy density is defined as
%
\begin{equation}
\label{Sth}
S=-\left(\partial \Omega /\partial T \right)_{\mu} /V,
\end{equation}
%
i.e. as the temperature derivative at the fixed chemical potential $\mu$ with
$V$ being the system's volume.  Due to the stationarity property of $\Omega$,
the only $T$-dependences to be taken into account in calculating the entropy
are the explicit dependence of the discrete Matsubara frequencies on $T$ or in
the occupations $n (\varepsilon )$ in the integral formulation
(\ref{Om-maz-cont}). From eqs (\ref{Om-maz-cont}) and (\ref{Sth}) one finds
%
\begin{equation}
\label{Scorn} 
S = -
\mbox{Tr}  \int \dpi{ p}\frac{\partial n(\varepsilon)}{\partial T}
\left[- 2\Im\;\ln\left[-\Gr^R (p_0+\ii 0,\vec p)\right]
- \Re\Gr^{R}\Gamma -A\Re\Se^{R}\right]
-\frac{\partial \Phi_T}{\partial T}
\end{equation} 
%
for the entropy density. With the help of the identity (\ref{nt}) we replace
${\partial n(\varepsilon)}/{\partial T}$ in eq. (\ref{Scorn}) and then perform
partial integration over $p_0$. Thus, we obtain
%
\begin{eqnarray}
\label{ssum} 
S &=&S_{\scr{loc}} + S_{\scr{mem}}, \quad\quad\mbox{where}\\ 
\label{S-dyn} 
S_{\scr{loc}} &=& -
\mbox{Tr} \int \dpi{ p}\sigma (\varepsilon)\frac{\partial}{\partial p_0}
\left[-2\Im\;\ln[-\Gr^R(p_0 +\ii 0 ,\vec p )]
- \Re\Gr^R\Gamma \right],\\ 
\label{scor2-th}
S_{\scr{mem}} &=&-
\frac{\partial \Phi }{\partial T} + \mbox{Tr} \int \dpi{ p}
\sigma (\varepsilon)
\frac{\partial\left(A\Re\Se^R\right)}{\partial p_0}, \\ 
\quad\sigma(\varepsilon)&=&\mp(1\mp n(\varepsilon))
\ln(1\mp n(\varepsilon))-n(\varepsilon)\ln n(\varepsilon).
\end{eqnarray}
%
We have used subscripts ``local'' and ``memory'' to denote these two different
contributions, because below we demonstrate that they are indeed associated
with local (Markovian) $s_{\scr{loc}}^0$, cf. eq. (\ref{S(gen)}), and memory
(non-Markovian) $s_{\scr{mem}}^0$, cf. eq. (\ref{entr-flow-Mark}), parts of
the kinetic entropy. Taking derivatives in eq. (\ref{S-dyn}), we readily get
%
\begin{equation}
\label{Sequil-zero} 
S_{\scr{loc}}  =\mbox{Tr}
\int \dpi{ p}\sigma (\varepsilon)
A^0_s (p)  
\end{equation}
%
with $A^0_s$ defined in eq. (\ref{As}). Thus, $S_{\scr{loc}}$, indeed,
coincides with 0-component of the kinetic entropy flow 
(\ref{entr-transp}).

In order to clarify the meaning of the values $S_{\scr{loc}}$ and
$S_{\scr{mem}}$ we first inspect the quasiparticle limit, in which the
spectral function is reduced to the delta-function, cf. eq.
(\ref{dyn-qp-lim}). In this limit, the value (\ref{Sequil-zero}) for
$S_{\scr{loc}}$ is reduced to
%
\begin{equation}
\label{Sqp} 
S_{\scr{loc}}^{\scr{qp}}=\mbox{Tr}\int \frac{\di^{3}p}{(2\pi)^3}
\sigma \left(\varepsilon (\vec p)-\mu\vphantom{Z^R}\right),
\end{equation}
%
where the quasiparticle energy $\varepsilon (\vec p)$ is determined by
solution of dispersion equation (\ref{qp-disp-non}). The full
$S_{\scr{loc}}^{\scr{qp}}$ is just the sum of single-particle contributions,
as if we deal with a noninteracting ideal gas of quasiparticles. This is a
standard picture in the quasiparticle approximation. Corrections to
$S_{\scr{loc}}^{\scr{qp}}$, resulting from $S_{\scr{loc}}$, are of higher
order in the width $\Gamma$.  At the same time, $S_{\scr{mem}}$ provides
corrections to $S_{\scr{loc}}^{\scr{qp}}$ even in the zero-order in $\Gamma$,
which are associated with real rescatterings of on-mass-shell quasiparticles.
This fact was demonstrated by Carneiro and Pethick in ref. \cite{Carneiro}.
For the Fermi liquid, they showed that the first diagram for $\Phi$ that
contributes to $S_{\scr{mem}}$ is the triangle one $\Phi^{(3)}$ of eq.
(\ref{Phi-123}). Contributions of $\Phi^{(1)}$ and $\Phi^{(2)}$ are zero. In
the quasiparticle approximation, the contribution of $\Phi^{(3)}$ is (cf. eqs
(45) and (72) of ref. \cite{Carneiro})
%
\begin{eqnarray}
\label{s-2-pethick}
S_{\scr{mem}}^{\scr{qp}}= \frac{1}{12} 
\int \dpi{p} \left(\Gamma_{\scr{b}}^{\scr{qp}} (p)\right)^3
\frac{\partial n_{\scr{b}} (p_0)}{\partial T}, 
\end{eqnarray}
%
where $n_{\scr{b}}$ is the thermal 
occupation number of artificial boson, given by eq. (\ref{occup}), 
and the bosonic width
$\Gamma_{\scr{b}}$ is defined as 
%
\begin{equation}\label{G-expl} 
\Gamma_{\scr{b}} (p)=
V_0 d
\int \dpi{ p'} \left[ 
n_{\scr{f}} \left(\varepsilon'-\frac{1}{2}\omega\right)
n_{\scr{f}} \left(\varepsilon'+\frac{1}{2}\omega\right)\right]
A_{\scr{f}}\left(p'-\frac{1}{2}p\right) A_{\scr{f}}\left(p'+\frac{1}{2}p\right) 
\end{equation}
%
with $\varepsilon' = p'_0 -\mu$ and $\omega=p_0$, $n_{\scr{f}}$ stands for
thermal fermion occupations, cf. eq.  (\ref{occup}). As above, $V_0$ stands
for the strength of the two-body potential. To get the quasiparticle
approximation to this width ($\Gamma_{\scr{b}}^{\scr{qp}}$), we should replace
the exact spectral function $A_{\scr{f}}$ in eq. (\ref{G-expl}) by its
quasiparticle approximation $A^{\scr{qp}}_{\scr{f}}$ defined by eq.
(\ref{dyn-qp-lim}). Note that $S_{\scr{mem}}^{\scr{qp}}$ is expressed in terms
of bosonic quantities $n_{\scr{b}}$ and $\Gamma_{\scr{b}}$, although we have
started initially with the purely fermionic system with potential interaction.
This fact provides a link to the thermodynamic calculation of Riedel
\cite{Riedel}, where the correction to the standard quasiparticle entropy of a
Fermi liquid is presented in the form of the effective bosonic contribution.
We remind that Riedel \cite{Riedel} summed all the series of ring diagrams
rather than considered only three first of them as in eq. (\ref{Phi-123}). At
low temperatures, $S_{\scr{mem}}^{\scr{qp}} \sim T^3 \ln T$ \cite{Carneiro},
i.e. it gives the leading correction to the standard quasiparticle entropy.
This is the famous correction to the specific heat of liquid $^3$He
\cite{Baym91,Riedel,Carneiro}. Since this correction is quite comparable
(numerically) to the leading term in the specific heat ($\sim T$), we may
claim that the liquid $^3$He is a liquid with quite strong memory from the
point of view of kinetics.

Please also notice that different but an even more simple form of
thermodynamic entropy including memory corrections follows from eq.
(\ref{Energy-Pressure}) if one uses thermodynamic relation $E+PV-\mu N=TSV$:
\begin{equation}\label{Sequil-dif}
TS = \mbox{Tr}  \int \dpi{p} 
A  n(\varepsilon)\left[ \varepsilon - \frac{2}{3}\varepsilon_p^{0}\right].
\end{equation}

\subsection{Non-Markovian Entropy in Equilibrium}

We now evaluate the memory correction $(s_{\scr{mem}}^{\nu})_{\scr{eq}}$ to
the kinetic entropy flow (see eq. (\ref{s3-Mark})) in thermal equilibrium.
Proceeding from the definition of $\Loop^{ij}$ (\ref{Loop-p}), as well as from
equilibrium relations (\ref{Geq})--(\ref{occup}), and identity
(\ref{fb-occup}), we present $\Loop^{ij}$ in the form
%
\begin{eqnarray}
\label{L(-+)eq} 
\Loop^{-+} (p) =
\ii n_{\scr{b}}(\omega) \Gamma_{\scr{b}} (p),\,\,\,\,
\Loop^{+-} (p) =
\ii \left[1+n_{\scr{b}}(\omega)\right] \Gamma_{\scr{b}} (p), 
\end{eqnarray}
%
where $\omega = p_\nu U^\nu$, $\Gamma_{\scr{b}}$ is defined by eq.
(\ref{G-expl}), and $n_{\scr{b}}$ is the bosonic occupation number. Now,
$(s_{\scr{mem}}^{\nu})_{\scr{eq}}$ of eq. (\ref{s3-Mark}) takes the form
%
\begin{eqnarray}
\label{Ieq-small} 
(s_{\scr{mem}}^{\nu})_{\scr{eq}} &=& -
\frac{1}{2}  \int \dpi{p} 
n_{\scr{b}}(\omega) \left[1+n_{\scr{b}}(\omega)\right]
\Gamma_{\scr{b}}^2 (p) \left(\frac{\omega}{T}-1\right)
\frac{\partial }{\partial p_\nu} 
\left[n_{\scr{b}}(\omega)\Gamma_{\scr{b}} (p)\right]\\ 
\label{Dsmall-1} 
&=& -
\frac{1}{2} \int \dpi{p} 
n_{\scr{b}} \left(1+n_{\scr{b}}\right)
\left(\frac{\omega}{T}-1\right)
\left(\Gamma_{\scr{b}}^3 \frac{\partial
n_{\scr{b}}}{\partial p_\nu}
+\frac{1}{3}\frac{\partial \Gamma_{\scr{b}}^3}{\partial p_\nu}n_{\scr{b}}\right)  
\nonumber\\
&=& 
\frac{1}{6}  U^\nu\int \dpi{p} 
\Gamma_{\scr{b}}^3 
\left[ 
n_{\scr{b}}^2 \left(1+n_{\scr{b}}\right)
\frac{\omega}{T^2}\right]. 
\end{eqnarray}
%
The last line is obtained with the help of the partial integration, explicitly
taking derivatives of bosonic occupations and using their property $\di
n_{\scr{b}}/\di \omega = - n_{\scr{b}} (1+n_{\scr{b}})/T$.  Now, we change the
integration variable $p \to -p$ in the last line of eq. (\ref{Dsmall-1}) and
use the parity properties of the relevant quantities
%
\begin{eqnarray}
\label{gamsmall}
\Gamma_{\scr{b}} (-\omega ) = -\Gamma_{\scr{b}} (\omega ),\,\,\,\, 
n_{\scr{b}} (-\omega ) = - \left[1+n_{\scr{b}} (\omega ) \right],
\end{eqnarray}
%
and arrive at 
%
\begin{eqnarray}
\label{Dsmall-2} 
(s_{\scr{mem}}^{\nu})_{\scr{eq}} &=& -
\frac{1}{6}   U^\nu\int \dpi{p} 
\Gamma_{\scr{b}}^3 
\left[ 
n_{\scr{b}} \left(1+n_{\scr{b}}\right)^2
\frac{\omega}{T^2}\right]. 
\end{eqnarray}
%
Taking the half-sum of the r.h.s. in (\ref{Dsmall-2}) and the last line of eq.
(\ref{Dsmall-1}) and using the identity (\ref{nbt-ident}) we finally arrive at
%
\begin{eqnarray}
\label{Dsmall} 
(s_{\scr{mem}}^{\nu})_{\scr{eq}} &=& 
\frac{1}{12}  U^\nu \int \dpi{p} 
\Gamma_{\scr{b}}^3 
\frac{\partial n_{\scr{b}}}{\partial T}.  
\end{eqnarray}
%
The 0-component of (\ref{Dsmall}) precisely coincides with the
thermodynamic
quantity
$S_{\scr{mem}}^{\scr{qp}}$ given by eq. (\ref{s-2-pethick}) in the quasiparticle
approximation, provided we consider it in the rest frame of the matter, i.e.
at $U^\nu=\{1,\vec 0\}$ . This fact again justifies the name ``memory'' for
the thermodynamic quantity $S_{\scr{mem}}$.

Thus, we have demonstrated that our kinetic entropy, including memory
contributions, coincides with the thermodynamic entropy in thermal
equilibrium.

\section{Conclusion and Prospects}

Generalized kinetic equations have been derived from the Kadanoff-Baym
equations by means of the gradient approximation. Thereby the usual restriction
to small mass widths (mass-shell condition) for the particles involved is
avoided. The equations are valid, if all phase-space distribution functions
vary slowly across the space-time region.  Since the mass-shell relation
between energy and momentum is lost for unstable or virtual particles, the
generalized kinetic equation is formulated in terms of a distribution function
depending on energy and spatial momentum.  Mass-shell equation loses its
normal sense in context of quasiparticle approximation and rather
becomes equivalent to the generalized kinetic equation.
Alongside a local retarded equation
has to be solved which provides the dynamical information about the spectral
functions of the particles.  Besides the usual drift and collision terms
present in any transport equation, like in Landau's Fermi liquid theory
\cite{Land56,Abrikos,Baym91}, a genuine width and a 
fluctuation-dependent term appear in the generalized transport 
equation.  The latter term is normally  dropped along with the 
quasiparticle approximation which we avoided.  This extra term 
gives rise to backflow contribution in the current and 
produces width and fluctuation dependent contributions to the 
energy momentum tensor pertaining to this transport scheme. It 
is indeed essential in order to preserve the conservation laws 
in the case of broad damping widths.

We suggest to follow Baym's $\Phi$-derivable principle to consistently
construct all terms for both, the kinetic and the retarded equation. It has a
couple of important advantages, cf. ref. \cite{Baym}. First, it leads to
closed, i.e. self-consistent equations which can be closed at any order or
loop level of the diagrams of $\Phi$, this way defining an effective theory.
We showed in ref. \cite{IKV} that the original properties of $\Phi$ are also
valid for genuine nonequilibrium systems described within the real-time
formalism, namely that the so constructed approximation is conserving and at
the same time thermodynamically consistent.  In fact, for this to hold Baym
showed that the $\Phi$-derivable scheme is not only sufficient but even
necessary!  The $\Phi$-derivable energy momentum tensor has explicitly been
constructed for the contour Dyson equation in ref. \cite{IKV}. Here we showed
that the conserving properties also hold for the here derived generalized
kinetic equations.  As further shown in ref. \cite{IKV} the scheme can be
easily extended to include classical boson fields dynamically, such as mean
fields or condensates. The latter permit to include soft modes in terms of
these classical fields, much in the spirit of the kinetic picture
\cite{Blaizot,Jackiw93,Knoll95} of hard loop approximations
\cite{BraatenPisarski}.

The structure of the collision term was studied by means of representing the
$\Phi$ functional in terms of ``$-+$'' and ``$+-$'' Green's functions which
represent the Wigner phase-space densities \cite{Knoll95}. The advantage of
the ``$-+$'' and ``$+-$'' representation is that it leads to a natural
decomposition of the collision term into multi-particle processes with Feynman
transition amplitudes which determine the partial rates. 
Furthermore, it has been discussed that sometimes it appears
advantageous to account for such memory effects by including new ``artificial''
particles (e.g.  bosonization of particle-hole excitations) which then lead to
local collision terms.

We also addressed the question whether a closed
nonequilibrium system approaches the thermodynamic equilibrium during its
evolution. Investigating the structure of the collision term in the
$\Phi$-derivable scheme we obtained definite expressions for the local
(Markovian) entropy flow and were able to explicitly demonstrate the
$H$-theorem for some of the common choices of $\Phi$ approximations.  The
expression for the local entropy flow holds beyond the quasiparticle picture,
and thus generalizes the well-known Boltzmann expression for the kinetic
entropy.  To demonstrate memory effects in the
generalized kinetics, we considered a particular case of a system of fermions
interacting via two-body zero-range potential.  We calculated the memory
(non-Markovian) contribution to the kinetic entropy, which merges the
equilibrium limit with its famous correction to the specific heat of liquid
$^3$He \cite{Riedel,Carneiro,Baym91}: $\sim T^3 \ln T$.  Since this correction
is quite comparable (numerically) to the leading term in the specific heat
($\sim T$), one may claim that the liquid $^3$He is a liquid with quite strong
memory effects from the point of view of kinetics.

The mass-width effects are important for description of various physical
systems.  As for immediate applications of the developed formalism, we see the
description of wide resonances (such as $\rho$-meson, $\Delta$-resonance,
etc.) in nonequilibrium hadron matter produced in heavy-ion collisions. Since
the widths of these resonances are of the order (or even larger) than the
excitation energy in the system, a self-consistent treatment of these widths
is required.  Up to now, width effects were considered either within some
simplified dynamics with phenomenological Landau--Migdal residual interaction
\cite{MSTV,Vos93} or within a simple $\Phi$-derivable approximation at thermal
equilibrium and in the dilute gas limit \cite{Weinhold}. In particular, it was
demonstrated in ref. \cite{Knoll95}, that the soft-photon production is
sensitive to dynamical and width effects. Interplay between the width and
in-medium population of $\rho$-mesonic states may also simulate a shift of the
$\rho$-meson mass in the nuclear medium, and thus affects the production of
di-leptons in relativistic heavy-ion collisions.  It is of interest to study
these effects within a dynamical approach, such as the scheme presented here.

Further applications concern relativistic plasmas, such as QCD and QED
plasmas. The plasma of deconfined quarks and gluons was present in the early
Universe, it may exist in cores of massive neutron stars, and may also be
produced in laboratory in ultra-relativistic nucleus--nucleus collisions.  All
these systems need a proper treatment of particle transport.  Perturbative
description of soft-quanta propagation suffers of infrared divergences and one
needs a systematic study of the particle mass-width effects in order to treat
them, cf. ref. \cite{Knoll95}.  A thermodynamic $\Phi$-derivable approximation
for hot relativistic QED plasmas---a gas of electrons and positrons in a
thermal bath of photons---was recently considered by Vanderheyden and Baym
\cite{Vanderheyden}. Their treatment may be also applied to the
high-temperature super-conductors and the fractional quantum Hall effect
\cite{Halp93,Ioff89}. Our approach allows for a natural generalization of such
a $\Phi$-derivable schemes to the dynamical case.

Another application, as we see it, concerns the description of the neutrino
transport in supernovas and hot neutron stars during first few minutes of
their evolution. At an initial stage, neutrinos typically of thermal energy,
produced outside (in the mantel) and inside the neutron-star core, are trapped
within the regions of production.  However, coherent effects in neutrino
production and their rescattering on nucleons \cite{Knoll95} reduce the
opacity of the nuclear-medium and may allow for soft neutrinos to escape the
core and contribute to the heating off the mantle. The extra energy transport
may be sufficient to blow off the supernova's mantle in the framework of
the shock-reheating mechanism \cite{Wilson}. The description of the neutrinos
transport in the semi-transparent region should therefore be treated with the
due account of mass-widths effects.

\section*{Acknowledgments}

We are grateful to G.E. Brown, P. Danielewicz, H. Feldmeier, B. Friman, H. van
Hees, and E.E. Kolomeitsev for fruitful discussions and suggestions.  Two of
us (Y.B.I. and D.N.V.) highly appreciate the hospitality and support rendered
to us at Gesellschaft f\"ur Schwerionenforschung and thank the Niels Bohr
Institute for its hospitality and partial support during the completion of
this work.  This work has been supported in part by BMBF under the program on
scientific-technological collaboration (WTZ project RUS-656-96).

\section*{APPENDICES}
\appendix
\section{Matrix Notation} \label{Contour}

In calculations that apply the Wigner transformations, it is necessary to
decompose the full contour into its two branches---the {\em time-ordered} and
{\em anti-time-ordered} branches. One then has to distinguish between the
physical space-time coordinates $x,\dots$ and the corresponding contour
coordinates $x^{\cal C}$ which for a given $x$ take two values
$x^-=(x^-_{\mu})$ and $x^+=(x^+_{\mu})$ ($\mu\in\{0,1,2,3\}$) on the two
branches of the contour (see figure 1).  Closed real-time contour integrations
can then be decomposed as
%
\begin{eqnarray}
\label{C-int}
\oint\di x^{\cal C} \dots =\int_{t_0}^{\infty}\di x^-\dots
+\int^{t_0}_{\infty}\di x^+\dots
=\int_{t_0}^{\infty}\di x^-\dots -\int_{t_0}^{\infty}\di x^+\dots, 
\end{eqnarray}
%
where only the time limits are explicitly given.  The extra minus sign of the
anti-time-ordered branch can conveniently be formulated by a $\{-+\}$
''metric'' with the metric tensor in $\{-+\}$ indices
%
\begin{eqnarray}
\label{sig}
\left(\sigma^{ij}\right)&=&
\left(\sigma_{ij}\vphantom{\sigma^{ij}}\right)=
{\footnotesize\left(\begin{array}{cc}1&0\\ 
0& -1\end{array}\right)}
\end{eqnarray}
%
which provides a proper matrix algebra for multi-point functions on the
contour with ''co''- and ''contra''-contour values.  Thus, for any two-point
function $F$, the contour values are defined as
%
\begin{eqnarray}\label{Fij}
F^{ij}(x,y)&:=&F(x^i,y^j), \quad i,j\in\{-,+\},\quad\mbox{with}\cr
F_i^{~j}(x,y)&:=&\sigma_{ik}F^{kj}(x,y),\quad
F^i_{~j}(x,y):=F^{ik}(x,y)\sigma_{ki}\cr
F_{ij}(x,y)&:=&\sigma_{ik}\sigma_{jl}F^{kl}(x,y),
\quad\sigma_i^k=\delta_{ik}
\end{eqnarray}
%
on the different branches of the contour. Here summation over repeated indices
is implied. Then contour folding of contour two-point functions, e.g. in Dyson
equations, simply becomes
%
\begin{eqnarray}\label{H=FG}
H(x^i,y^k)=H^{ik}(x,y)=\oint\di z^{\cal C} F(x^i,z^{\cal C})G(z^{\cal C},y^k)
=\int\di z F^i_{~j}(x,z)G^{jk}(z,y)
\end{eqnarray}
%
in the matrix notation.

For any multi-point function the external point $x_{max}$, which has the
largest physical time, can be placed on either branch of the contour without
changing the value, since the contour-time evolution from $x_{max}^-$ to
$x_{max}^+$ provides unity. Therefore, one-point functions have the same value
on both sides on the contour.

Due to the change of operator ordering, genuine multi-point functions are, in
general, discontinuous, when two contour coordinates become identical. In
particular, two-point functions like $\ii F(x,y)=\left<\Tc {\widehat
    A(x)}\medhat{B}(y)\right>$ become
%
\begin{eqnarray}\label{Fxy}
\hspace*{-0.5cm}\ii F(x,y) &=&
\left(\begin{array}{ccc} 
\ii F^{--}(x,y)&&\ii F^{-+}(x,y)\\[3mm]
\ii F^{+-}(x,y)&&\ii F^{++}(x,y)
\end{array}\right)=
\left(\begin{array}{ccc} 
\left<{\cal T}\medhat{A}(x)\medhat{B}(y)\right>&\hspace*{5mm}&
\mp \left<\medhat{B}(y)\medhat{A}(x)\right>\\[5mm]
\left<\medhat{A}(x)\medhat{B}(y)\right>
&&\left<{\cal T}^{-1}\medhat{A}(x)\medhat{B}(y)\right>
\end{array}\right), 
\end{eqnarray}
%
where ${\cal T}$ and ${\cal T}^{-1}$ are the usual time and anti-time ordering
operators.  Since there are altogether only two possible orderings of the two
operators, in fact given by the Wightman functions $F^{-+}$ and $F^{+-}$,
which are both continuous, not all four components of $F$ are independent. Eq.
(\ref{Fxy}) implies the following relations between nonequilibrium and usual
retarded and advanced functions
%
\begin{eqnarray}\label{Fretarded}
F^R(x,y)&=&F^{--}(x,y)-F^{-+}(x,y)=F^{+-}(x,y)-F^{++}(x,y)\nonumber\\
&:=&\Theta(x_0-y_0)\left(F^{+-}(x,y)-F^{-+}(x,y)\right),\nonumber\\
F^A (x,y)&=&F^{--}(x,y)-F^{+-}(x,y)=F^{-+}(x,y)-F^{++}(x,y)\nonumber\\
&:=&-\Theta(y_0-x_0)\left(F^{+-}(x,y)-F^{-+}(x,y)\right),
\end{eqnarray}
%
where $\Theta(x_0-y_0)$ is the step function of the time difference.  The
rules for the co-contour functions $F_{--}$ etc. follow from eq. (\ref{Fij}).

For such two point functions complex conjugation implies
%
\begin{eqnarray}\label{ComplexConjugate}
\left(\ii F^{-+}(x,y)\right)^*&=&\ii F^{-+}(y,x)
\quad\Rightarrow\quad \ii F^{-+}(X,p)=\mbox{real},\nonumber\\
\left(\ii F^{+-}(x,y)\right)^*&=&\ii F^{+-}(y,x)
\quad\Rightarrow\quad \ii F^{+-}(X,p)=\mbox{real},\nonumber\\
\left(\ii F^{--}(x,y)\right)^*&=&\ii F^{++}(y,x)
\quad\Rightarrow\quad \left(\ii F^{--}(X,p)\right)^*=\ii F^{++}(X,p),\nonumber\\
\left(F^R(x,y)\right)^*&=&F^A(y,x)
\quad\hspace*{3.5mm}\Rightarrow\quad \left(F^R(X,p)\right)^*=F^A(X,p),
\end{eqnarray}
%
where the right parts specify the corresponding properties in the Wigner
representation. Diagrammatically these rules imply the simultaneous swapping
of all $+$ vertices into $-$ vertices and vice versa together with reversing
the line arrow-sense of all propagator lines in the diagram.

In components the determination of the self-energy from the functional
variation of $\Phi$ (cf. eq. (\ref{varphi})) reads
%
\begin{eqnarray}\label{var-Phi-component}
-\ii 
\Se_{ik} (x,y)=\mp\frac{\delta\ii\Phi}{\delta\ii \Ga^{ki} (y,x)}
\quad\Rightarrow\quad
-\ii 
\Se_{ik}(X,p)=\mp\frac{\delta\ii\Phi}{\delta\ii \Ga^{ki}(X,p)},
\quad i,k\in\{-+\}
\end{eqnarray}
%
the right expression given in the Wigner representation. Note that the
variation over a $\{+-\}$-``contra-variant'' Green's function $\Ga^{ki}$
produces a $\{+-\}$-``covariant'' self-energy $\Se_{ik}$, cf. (\ref{Fij}).
This occurs due to the same reason discussed above, cf. eq.  (\ref{C-int}),
when dealing with $\{+-\}$ matrix notation with integrations over physical
times, rather than contour times.  The extra minus signs occurring for the
anti-time ordered branches are precisely taken into account by the
``covariant'' notation (\ref{Fij}).

\section{Equilibrium Relations}\label{eq.rel}

For completeness of the thermodynamic consideration, we explicitly present
here equilibrium relations between quantities on the real-time contour.
Basically, they follow from the Kubo--Martin--Schwinger condition \cite{Kubo}
%
\begin{eqnarray}
\label{KMS-G}
\Ga^{-+}(p) = \mp\Ga^{+-}(p) e^{-\varepsilon/T}, \,\,\,\,
%
%
\Sa^{-+}(p) = \mp\Sa^{+-}(p) e^{-\varepsilon/T}, 
\end{eqnarray}
%
where $\varepsilon = p_\nu U^\nu - \mu$ with $U^\nu$ and $\mu$ being a global
4-velocity of the system and a chemical potential related to the charge,
respectively. All the Green's functions can be expressed through the retarded
and advanced Green's functions
%
\begin{eqnarray}
\label{Geq}
\Ga^{ik}(p) = 
\left(\begin{array}{ccc}       
\left[1\mp n(\varepsilon)\right]\Ga^R(p)\pm n(\varepsilon) \Ga^A(p) &&
\pm \ii n(\varepsilon) \A(p)\\[3mm]  
-\ii \left[1\mp n(\varepsilon)\right] \A(p) &&
-\left[1\mp n(\varepsilon)\right] \Ga^A(p)\mp n(\varepsilon) \Ga^R(p)
\end{array}\right), 
\end{eqnarray}
%
$i,k$ mean $+$ or $-$, and the self-energies take a similar form 
%
\begin{eqnarray}
\label{Seq}
\Sa_{ik}(p) = 
\left(\begin{array}{ccc}       
\Sa^R(p) \pm \ii n(\varepsilon) \Gm(p) && 
\mp\ii n(\varepsilon) \Gm(p) \\[3mm]
\ii \left[1\mp n(\varepsilon)\right]\Gm(p)&&
- \Sa^A(p) \pm \ii n(\varepsilon) \Gm(p) 
\end{array}\right). 
\end{eqnarray}
%
Here 
%
\begin{equation}\label{occup}
n(\varepsilon) =\left[\exp(\varepsilon/T)\pm 1\right]^{-1}\;  
\end{equation}
%
are thermal Fermi/Bose--Einstein occupations. They obey some useful relations
between fermion $n_{\scr{f}}$ and boson $n_{\scr{b}}$ occupation numbers, like
%
\begin{equation}\label{fb-occup}
n_{\scr{f},\scr{b}}(\varepsilon + \omega/2)
\left[1 \mp n_{\scr{f},\scr{b}}(\varepsilon - \omega/2)\right]
=
\left[ n_{\scr{f},\scr{b}}(\varepsilon -\omega/2)
-
n_{\scr{f},\scr{b}}(\varepsilon + \omega/2) \right]
n_{\scr{b}}(\omega),
\end{equation}
%
or derivatives with respect to $T$
%
\begin{eqnarray}
\label{nt} 
\frac{\partial n (\varepsilon)}{\partial T}&=&-
\frac{\partial \sigma (\varepsilon)}{\partial \varepsilon} ,\,\,\,\,
\sigma (\varepsilon ) = \mp [1\mp n (\varepsilon )]
\ln[1\mp n (\varepsilon)]-n (\varepsilon)\ln n (\varepsilon),\\
\label{nbt-ident}
\frac{\partial n (\varepsilon)}{\partial T}&=& 
- \frac{\omega}{T^2} n_{\scr{b}}
\left(1\mp n_{\scr{b}}\right). 
\end{eqnarray}
%
The link between the Matsubara technique and the real-time formulation used
here can be provided by extending the real time contour by an imaginary tail
going to $-\ii\beta$, $\beta=1/T$, this way defining the equilibrium density
operator $\exp(-\beta \medhat{H})$. Thus, the link is provided by 
considering analytic expressions like the
''contour trace'' of two-point functions $\ii F_{eq}(x,y)=\left<\Tc
  {\medhat{A}(x)}\medhat{B}(y)\right>$
%
\begin{eqnarray}\label{Real-Mats}
\int_{{\cal C}\{{\rm eq}\}}F_{\rm eq}(t,t+0)\di t
&=&\int_0^{-\ii\beta}F_{\rm eq}^{-+}(t,t)\di t 
=\mp\ii\beta\int_{-\infty}^{\infty}
\frac{\di\omega}{2\pi}\; n(\varepsilon)
\left(F^{A}_{\rm eq}(\omega)-F^{R}_{eq}(\omega)\right)\nonumber\\
&=&\sum_{m=-\infty}^{\infty}\int_{-\infty}^{\infty}\frac{\di\omega}{2\pi}
\;\frac{F^{S}_{\rm eq}(\omega)}{\ii\varepsilon_m+\mu_{AB}-\omega}
=\sum_{m=-\infty}^{\infty} F^{Matsubara}(\ii\varepsilon_m).  
\end{eqnarray}
%
Here the contour time $t+0$ is placed infinitesimally behind $t$ on the
contour in order to specify a {\em fixed} operator ordering of the two
external operators of $F$, $\mu_{AB}$ is the chemical potential associated to 
$\left<{\medhat{A}(x)}\medhat{B}(y)\right>$, and  $\mu_{AB}=-\mu_{BA}$. 
The step towards the discrete Matsubara sum is
provided by standard residue technique, cf. ref. \cite {FetterWalecka}, fig.
25.4 ff. The sum runs over the Matsubara energies
%
\begin{eqnarray}\label{Mats-energy}
\varepsilon_m=\left\{\begin{array}{ll}
(2m+1)\pi T\quad&\mbox{\rm for fermions}\\
2m\pi T\quad&\mbox{\rm for bosons.}\end{array}\right. 
\end{eqnarray}
%
Thereby, the Matsubara form of the two-point function $F_{eq}$ has the
spectral representation
%
\begin{eqnarray}\label{Mats-spectral}
F^{Matsubara}(z)=\int_{-\infty}^{\infty}
\frac{\di\omega}{2\pi}\;\frac{F^{S}_{\rm eq}(\omega)}{z+\mu_{AB}-\omega}
=\left\{
\begin{array}{ll}
    F^{R}_{\rm eq}(z+\mu_{AB})\quad&\mbox{for Im $z>0$}\\[3mm]
    F^{A}_{\rm eq}(z+\mu_{AB})\quad&\mbox{for Im $z<0$}
\end{array}\right.
\end{eqnarray}
%
in terms of the {\em real time contour} spectral function $F^{S}_{\rm
  eq}=-2\Im F^{R}_{\rm eq}$.

\section{Diagram rules}\label{diagrules}

For relativistic theories with local vertices the diagrammatic rules on the
contour are identical to the standard Feynman rules except that all time
integrations are to be replaced by contour integrations. The diagrams
contributing to $\Phi$ are calculated as diagrams with one external
point, namely the interaction part of the Lagrangian $\left<\Lint(x)\right>$,
then contour integrated over $x$ and weighted with $1/n_\lambda$, where
$n_\lambda$ counts the number vertices in the diagram. These diagrams have to
be two-particle irreducible with all lines representing full propagators. For
details about the corresponding diagrammatic rules on the contour see ref.
\cite{IKV}.

The rules for nonrelativistic two-body interactions are also naturally
extended to the contour $\cal C$ with
\begin{eqnarray}\label{Hint-non-rel}
\Hint(t_1)&=&{{\scriptsize\frac{1}{2}}}\int\di^3 x_1\oint\di^4 x_2
\medhat{\Psi}^\dagger({x}_1)
\medhat{\Psi}^\dagger({x}_2)
V({x}_1-{x}_2)
\medhat{\Psi}({x}_2)
\medhat{\Psi}({x}_1)\\
&&\mbox{with}\quad 
V(x_1-x_2)=U({\vec x}_1-{\vec x}_2)\;\delta_{\cal C}(t_1-t_2)
\end{eqnarray}
\def\Boson#1#2{\vphantom{\sum}
  \begin{picture}(3,.6)\thicklines\put(0.375,0.2){
           \multiput(0.1875,0)(.75,0){3}{\oval(.375,.375)[b]}
           \multiput(0.5625,0)(.75,0){3}{\oval(.375,.375)[t]}
           \put(0,0){\circle*{0.2}}\put(0,0.6){\makebox(0,0){$#1$}}
           \put(2.25,0){\circle*{0.2}}\put(2.25,0.6){\makebox(0,0){$#2$}}}
  \end{picture}
  }\unitlength0.5cm 
now defined for contour times $t_1$, $t_2$. 
One has however to observe that for the
instantaneous two-body interactions the diagrammatic elements are given by
$\Boson{}{}=-\ii V(x_1-x_2)$, i.e.  without the factor $1/2$ contained in
(\ref{Hint-non-rel}). With respect to the topological rules, indeed, the
interaction lines are treated like bosons in the relativistic theory with
local vertices. To this extend we recall that the expectation value for
$\left<\Hint(t)\right>=\frac{1}{2}\times\{\mbox{\it diagrams}\}$ has an
explicit factor $\frac{1}{2}$ in front of the corresponding diagrams.
Correspondingly, for the $n_\lambda$ vertex counting for the diagrams of $\Phi$
each two-body interaction line counts as {\em two} vertices, cf. the rules
given in ref. \cite{Abrikos} for the thermodynamic potential in Matsubara
formalism. However, the two-particle irreducibility of $\Phi$-diagrams is
defined with respect to cutting dynamical propagators only, i.e. leaving the
$V$-interaction lines untouched. In $\{-+\}$-matrix notation the above rules imply,
cf. \cite{Lif81} sect.  X,
\begin{eqnarray}\label{V-diag}
\left(\Boson++\right)^*&=&\Boson--=-\ii V(x_1-x_2)\nonumber\\[3mm]
\Boson-+&=&\Boson+-=0.
\end{eqnarray}

\end{document}